# Modelling Intra-driver Behavioral Adaptation through Risk Sensitivity and Regime Transitions: A Task-difficulty Car-following Model

Mohammad Tamim Kashifi

School of Architecture, Building, and Civil Engineering, Loughborough University, Leicestershire LE11 3TU, United Kingdom
Email: m.t.kashifi@lboro.ac.uk

**Abstract:** Over the past decade, there has been a growing trend toward integrating human factors (HF) into traffic flow models to better understand the complexities of human behavior and its impact on traffic dynamics. This research seeks to advance this trend by bridging the gap between traditional car-following models and the inherent variability of human driving behavior. By incorporating these elements, our models provide valuable insights into how driver behavior adaptation and risk-taking influence traffic flow dynamics. Specifically, the study proposes a model called Intelligent Driver Model Task Saturation that integrates human behavioral adaptations and risk-taking strategies into a modified version of the established Intelligent Driver Model, enriched by a cognitive layer based on Fuller's Task Capability Interface model. This amalgamation offers a perspective on the interplay between driver behavior adaptation when the driving task saturates and the risk-taking strategy of drivers. When total task demand exceeds task capacity, drivers may adapt their behaviors in accordance with Fuller's risk allostasis theory, aiming to mitigate risk to acceptable levels. The model was calibrated utilizing data from driving simulator scenarios involving both normal and distracted driving and naturalistic data, ensuring its sensitivity to diverse driving behaviors and adaptive behavioral changes. We investigated the model in terms of behavioral soundness and model fitting. Our results demonstrate that the model effectively incorporates endogenous mechanisms to explain both inter- and intra-driver heterogeneities in driving behavior. Additionally, it generates two plausible HF: risk-taking and behavior adaptation. The findings of this study suggest a step forward in achieving a more realistic representation of driving behavior adaptation and risk-taking strategies.

1. Introduction

Over the past decade, there has been a growing movement in the traffic flow theory community to integrate human factors (HF) into mathematical models of driving behavior (Saifuzzaman and Zheng, 2014). This reflects an increasing recognition that cognition, perception, and decision-making processes substantially influence a driver's ability to maintain awareness and make safe maneuvers. Incorporating such factors enables the development of more realistic models that respect psychological mechanisms underlying longitudinal dynamics control.

While it may appear that considering HF in traffic flow modeling is a recent trend, in reality, basic HF concepts have always been a fundamental aspect of traffic flow modeling. Over time, different schools of thought have developed, each distinguished by their unique assumptions about driver behavior, including how drivers react to various stimuli and which stimuli are most influential. Since the seminal work of Pipes (1953), a wide range of model formulations have been proposed. Broadly, there are five different types of car-following models: stimulus-based, safe headway, desired measurements, optimal velocity, and psycho-physical models (Zhu et al., 2018). The stimulus-based model's basic assumption is that the driver's response to the position and speed variations of the car in front governs the acceleration (deceleration) of the following vehicle (May, 1990). General Motors developed the most famous stimulus-based models. The second type of car-following model (i.e., safe headway) is built on the assumption that the following vehicle will select a speed and retain a spacing that the leading vehicle brakes suddenly, and the following vehicle could be brought to a safe stop. The safe headway theory is the basis of the Gipps model (1981). On the other hand, the desired measures techniques presuppose that a specific criterion for car-following, e.g., desired speed, is preferred by the driver of the following vehicle and continuously tries to maintain the differences in that measure to a minimum. The intelligent driver model (IDM), developed by Treiber et al. (2000), is based on this method. The optimum velocity models presuppose that every follower vehicle has an ideal safe speed that changes with the lead vehicle's position. The real velocity and the optimum velocity differences are used to calculate the following vehicle's acceleration (Bando et al., 1995). The fifth type of car-following modeling is based on psycho-physical theories. According to these theories, a driver's response will change, corresponding to the traffic situations in which they are involved. The relative speed and distance to the lead vehicle are often used to represent the boundary characteristics distinguishing the various states (Wiedemann, 1974).

Despite their historical contributions, there is a debate about these models' ability to represent human behavior convincingly. One of the limitations of these models is that they often fail to consider the individual characteristics and limitations of human drivers (Saifuzzaman and Zheng, 2014). More specifically, these engineering models (as put by Saifuzzaman and Zheng, 2014) are premised on the strict assumption of unlimited cognitive processing capacity. This assumption is markedly unrealistic, as it fails to accommodate the inherent limitations in drivers' task processing capacity. To illustrate, in relation to longitudinal driving behavior, drivers will adapt their driving behavior to compensate for any increase in task demand as the driving task escalates in difficulty. This adaptation can be as an increase in headway and/or reduction in speed. Such behavioral adaptation is not merely a reaction, but it is a fundamental principle embodied in Fuller's Task Capability Interface (TCI) model (Fuller, 2011). This model sheds light on the dynamics between the driver's capabilities and the demands of the driving task, highlighting the necessity for a more nuanced understanding of driving behavior, especially under increased task load.

To further explore the limitations of conventional models, several advancements have been made in modeling human driving dynamics, particularly with the inclusion of HF. For example, Yang and Peng (2010) introduced an error-able car-following model. The model determines the preferred acceleration for the driver by considering factors such as following distance, speed difference, and time headway. It also acknowledges the presence of uncertainties in the acceleration calculation. Specifically, the model considers that the driver's perception is less accurate when the following distance is considerable, allowing for greater potential deviation. Hamdar et al. (2015) modeled car-following behavior in congested situations. The risk-taking attitude of drivers was endogenously incorporated into the model using prospect theory. Recently, Kochi et al. (2023) investigated the influence of driving experience on drivers' perceptions and behaviors to prepare for traffic conflicts. The results showed that the driver's perception of task difficulty correlated with their driving speed, and inexperienced drivers tended to underestimate task difficulty compared to experienced drivers. The task difficulty and the feeling of risk were strongly correlated regardless of experience, and the estimation of statistical risk differed depending on experience. Further, the subjective task difficulty and objective safety margin were strongly correlated for experienced participants.

To operationalize the TCI concept, Saifuzzaman et al. (2015) proposed a framework for modeling car-following behavior incorporating HF. They introduced the concept of driving task difficulty, which is determined by the dynamic interplay between the demands of the driving task and the capabilities of the driver. The notion of task difficulty draws inspiration from the TCI model, a renowned theory that elucidates the factors influencing driver decision-making processes (Saifuzzaman et al., 2017, 2015). In the framework proposed by Saifuzzaman et al. (2015), the task difficulty variable (TD) is applied exogenously to alter the desired spacing or acceleration terms, typically by multiplying these expressions with TD or its inverse (1/TD). While this adjustment introduces sensitivity to perceived workload, it remains a scaling factor rather than an emergent behavioral response. There is no feedback loop linking TD to the driver's evolving state or performance, meaning that behavioral adaptation, such as compensatory responses under distraction, is not captured. As a result, the model can simulate parameter scaling but not true behavioral adjustment. Moreover, because the scaling operates multiplicatively, it may generate non-linear artefacts and unrealistic acceleration or spacing profiles when TD departs significantly from unity.

In a related study, van Lint & Calvert (2018) introduced a generic framework based on TCI and endogenously incorporated heterogeneities in driver traits and cognitive capacity. In this framework, the driving task is modeled using multiple layers. The researchers introduced HF variables such as task demand, capacity, and awareness to quantify the information processing costs associated with driving tasks and potential distractions. These costs are represented using fundamental diagrams of task demand (FDTD). Furthermore, the framework considers drivers' ability to adapt their responses according to Fuller's risk allostasis theory (Fuller, 2005) to minimize risk to acceptable levels (Fuller, 2005). Although this study paves the path toward the development of more sophisticated human driver models with the potential impact of HF, many aspects of driving behavior in this framework are yet unknown, or some assumptions are yet to be validated using ground truth data. In particular, the proposed FDTD relationships have not yet been empirically verified, and the framework's input requirements, such as real-time measures of task load and driver awareness, are abstract, difficult to observe, and challenging to quantify for model calibration or simulation. Consequently, the framework's practical applicability to large-scale or data-driven calibration remains limited.

Despite these advancements, a critical gap persists in capturing behavioral adaptation and risk sensitivity within mathematically tractable and empirically verifiable car-following models. Existing frameworks often oversimplify driver cognition, relying on static representations of task difficulty or workload, and seldom explore how adaptation and risk-taking interact to influence traffic stability and capacity. Moreover, few models have been validated against both controlled distraction and naturalistic trajectory data, limiting confidence in their behavioral soundness and operational relevance.

To address these shortcomings, the present study introduces the Intelligent Driver Model with Task Saturation (IDMTS), a behaviorally grounded car-following model that embeds human-factor mechanisms directly into the control law. The model endogenizes behavioral adaptation through distinct driving regimes, free driving, car-following, and behavioral adaptation, derived from Fuller's risk allostasis theory, allowing driver behavior to evolve dynamically in response to changing workload. To achieve this, we have adapted well-established IDM, layering a cognitive dimension atop the foundational physics theories of the original model. This additional layer is involvedly designed to ensure a fine balance between engineering and cognitive aspects of driving behavior. Behavioral adaptation is endogenized in our model, it emerges from distinct regimes driven by task-saturation theory. This allows behaviour to evolve dynamically rather than being imposed externally. Task saturation triggers behaviour adaptation regime where the driving task becomes challenging. Together, these enhancements bridge the gap between classical car-following theory and cognitive behavioral modelling, yielding a mathematically tractable yet behaviorally realistic framework

The structure of this study is as follows: Section 2 introduces the TCI model and interprets it through key traffic flow variables. In Section 3, we present our car-following model, focusing on the mechanisms underlying driver behavior adaptation, risk-taking strategies, and their integration into the proposed model framework. Section 4 details the calibration process of both the IDM+ and our proposed model. Section 5 presents the study's results and provides an in-depth discussion of the findings. Finally, Section 6 concludes the study, summarizing key insights and implications for future research.

## 2. Task Capability Interface (TCI) model and car-following control task

TCI model explains driver decision-making by recognizing the inherent limitations of perception and vehicle control. It posits that drivers continuously regulate their behavior to ensure that task demands do not exceed their capabilities (Fuller, 2011). The model is grounded in the principle of *task difficulty homeostasis*, whereby drivers adjust control variables such as speed and headway to keep perceived difficulty within tolerable limits. When task demand increases, as in the case of distraction, drivers may increase headway to reduce difficulty; conversely, when the task feels too simple, such as driving on a straight road, they may increase speed to restore a desired level of challenge (Saifuzzaman et al., 2015). Task saturation therefore converges towards an acceptable range shaped by both task capacity and motivational factors. We define the following variables that are required for our model development following the definitions of (Van Lint and Calvert, 2018) as Table 1.

**Tabel 1**: Definition of HF

| Term | Definition |
| --- | --- |
| Nominal Task Capacity | The standard capacity of a typical driver to process information for safe and effective task performance. This capacity is represented as TC = 1 (or 100%) (Van Lint and Calvert, 2018). |
| Driver Task Capacity | The processing capacity of a specific driver, measured in terms of the standard task capacity. |
| Driver Task Demand | This refers to the amount of information processing needed by a specific driver to safely and efficiently perform a certain task (i.e., a), denoted in units of TC as $TD_i^a(t)$. |
| Total Task Demand | This represents the cumulative task demand from all tasks for a specific driver as Eq. (1).<br>$$TD_i(t) = \sum_a TD_i^a(t) \qquad (1)$$ |

| Driver Task Saturation $TS_i(t)$ | This is a metric that represents the ratio of the total task demand $TD_i(t)$ for a specific driver to their individual task capacity $TC_i(t)$. It indicates how saturated or occupied a driver is in relation to their capacity as Eq. (2). |
|---|---|
| | $$TS_i(t) = \frac{TD_i(t)}{TC_i(t)} \tag{2}$$ |

According to Fuller (2002), the task demand on a driver at any given moment can be influenced by the speed of their vehicle and the spacing from the vehicle ahead of them. Based on this, we would like to express task saturation in terms of traffic characteristics in the following text.

Assumptions I: *Drivers' task demand increases with an increase in speed*. The faster a vehicle travels, the less time a driver has to recognize a potential hazard, process that information, make a decision, and then execute a maneuver (e.g., braking or steering). This rapid sequence of events places a high demand on the driver's cognitive and physical resources.

Assumption II: *Drivers' task demand decreases with an increase in spacing*. The closer a driver is to the vehicle in front of them, the less time they have to react in case the preceding vehicle suddenly brakes or changes direction. This closeness, combined with the necessity for rapid reaction, amplifies the task demand on the driver (Saifuzzaman et al., 2015).

When either speed increases or spacing decreases (or both), the margin for error reduces. Consequently, there's less time available for comprehension, decision-making, and response execution. Other factors being constant, this results in a heightened task demand on the driver, emphasizing the need for heightened awareness and quicker reflexes. In simpler terms, driving at higher speeds can compound the complexity of driving tasks, making it more challenging and demanding for the driver.

Assumption III: *Drivers with higher task capacity selects shorter desired time gap under certain conditions*. Drivers select desired time gap and prefer to keep that desired time gap. This is the time available for driver when they want to respond to a critical situation (e.g., harsh braking of leader). The shorter desired time gap than the desired one makes the driving task difficult and risky (Lewis-Evans et al., 2010). Empirical studies suggest that drivers with greater cognitive or perceptual capability often accept shorter headways, perceiving them as manageable (Heino et al., 1996; Hoogendoorn et al., 2013). However, this relationship is not universal. The choice of time gap is also moderated by individual preferences, situational risk perception, and contextual factors such as traffic density and distraction ((Lewis-Evans et al., 2010). Thus, while higher task capacity may enable shorter headway choices, this does not imply that all high-capacity drivers will systematically reduce their desired time gap. Instead, the relationship should be modulated by risk-taking behavior of drivers, reflecting a tendency rather than a rule. We follow Saifuzzaman et al. (2015) to represent $TS_i(t)$ in terms of traffic flow variables because it corroborates our assumptions I-III, which can be written as Eq. (3). Overall, high speed saturates the driving task, while extended spacing reduces the task saturation. Furthermore, selecting a short time headway represents a decreased task saturation for the driver.

$$TS_i(t) = \left(\frac{v_i(t)\, T_i}{s_i(t)}\right) \tag{3}$$

Where $v_i(t)$, and $s_i(t)$ are the current speed and spacing of the driver $i$ at time $t$, and $T_i$ indicates the desired time headway of driver $i$.

3. **Model Characteristics**

We consider IDM + of Schakel et al. (2012) to embed HF as our proposed model. The IDM+ is an evolution of IDM (Treiber et al., 2000) that segregates the free-driving and car-following strategies, opting to calculate the minimum rather than superimpose them. This approach yields more realistic capacity values (Van Lint and Calvert, 2018). The response (i.e., acceleration) in IDM+ behavior is determined as Eq. (4).

$$\frac{dv}{dt} = a_i \cdot min\left[1 - \left(\frac{v_i(t)}{v_{oi}}\right)^4, 1 - \left(\frac{s_i^*(v,\Delta v)}{s_i(t)}\right)^2\right], \tag{4}$$

where
$$s_i^*(v, \Delta v) = s_i(t) + v_i(t)\, T_i + \frac{v_i(t)\Delta v_i(t)}{2\sqrt{a_i b_i}}. \tag{5}$$

Parameters of IDM+ encompass; $a_i$, representing the maximum acceleration; $b_i$, indicating the maximum comfortable deceleration; $v_{0i}$, signifying the desired speed; $T_i$, denoting the desired headway; $s_0$, representing the stopping distance; and $i$ characterizes a specific driver. The original IDM+ model is formulated on certain assumptions, including maintaining a safe distance from the vehicle ahead, aiming to drive at a desired speed, and favoring acceleration/deceleration within a comfort zone. Moreover, it considers kinematic factors like the squared relationship between braking distance and speed. The model stimuli are the speed of ego vehicle, the speed difference with the leading vehicle, and spacing with the leading vehicle.

### 3.1 Behavior adaptation

The fundamental idea underpinning the behavior adaptation is the "task difficulty homeostasis" theory. This theory posits that drivers consistently make instantaneous decisions to ensure that the perceived level of task saturation of the driving task remains within their acceptable limit. They achieve this by modulating control elements such as speed and headway (Fuller, 2011), which is called driver behavior adaptation.

The behavior adaptation effects emerge due to a dynamic interplay between the total task demands and the capacity of the driver. Therefore, when the driver perceives that the task saturation exceeds their tolerable limit, such as driving while being distracted by a cell phone call, they tend to reduce their speed to bring the task saturation back within their acceptable range. Conversely, when the driving task appears overly uncomplicated, as seen in driving along a straight highway in free-flow conditions, the driver may accelerate to intensify the challenge. The scope of task difficulty that the driver aims for is established by their perception of their own abilities and their desire to engage with a specific level of task saturation.

The driving parameters influencing the task demand are the driving speed and spacing to the leading vehicle (Ray Fuller, 2002; R Fuller, 2002). These two parameters influence the available time for decision making and response to stimuli that, in the end, influences the task demand. The higher speed and short spacing mean a short time available for decision meaning and response that leads to a higher task demand (Saifuzzaman et al., 2015). If the driver's task capacity exceeds the task demand, the driving task is straightforward. On the other hand, if task demand exceeds the driver's task capacity, loss of control may occur (Fuller, 2011).

Delving deeper into contemporary research, the original IDM might not offer a comprehensive understanding of behavior adaptation effects in the context of longitudinal driving behavior. This limitation stems from IDM's lack of incorporation of crucial HF (Hoogendoorn et al., 2013). This suggests that for a more accurate representation of driving behavior adaptations, especially in scenarios of driver distractions, models should integrate and reflect human behaviors and characteristics more explicitly.

### 3.2 Risk-taking strategy

Drivers' behavioral adaptation occurs when they sense a need to adapt their behavior, especially when faced with an oversaturated driving task. Interestingly, not all drivers respond in the same manner to these heightened driving tasks. While some proactively adapt their behavior, others remain unresponsive to the escalating risks, continuing their journey without any behavioral adaptations. This variation underscores the importance of acknowledging the diverse range of responses and factoring in the heterogeneity of driving behaviors.

Through our analysis, two predominant driving strategies can be identified in car-following situations. The first group, *risk-taking drivers*, tend not to adapt their behaviour even when the task demand exceeds their capacity. Their safety margins are narrow, often operating closer to the boundaries of risk. In contrast, *risk-averse drivers* adapt their behaviour when the task demand surpasses their capacity. They prioritise safety by creating larger buffers or headways, ensuring sufficient reaction time even under challenging conditions. The behaviour adaptation regime of our model is designed to capture these divergent strategies, providing a more nuanced representation of driver heterogeneity in task-taking behaviour.

### 3.3 Mathematical formulation

Building on the behavioral framework introduced earlier, we formalize the proposed IDM Tast Saturation (IDMTS) model by specifying the acceleration function as a regime-dependent control law. In this model, behavioral adaptation and risk taking are endogenized in our model, it emerges from distinct regimes driven by task-saturation theory. This allows behaviour to evolve dynamically rather than being imposed externally. Task saturation triggers behaviour adaptation regime when the driving task becomes challenging. The central idea is that drivers do not respond uniformly across all contexts but instead operate within three qualitatively distinct regimes. Accordingly, acceleration is expressed as a function of spacing $s$, speed $v$, and relative speed $\Delta v$ as $\left(\frac{dv}{dt}\right)_{FDR} = f(s, v, \Delta v)$, where the functional form of $f(\cdot)$ depends on the prevailing regime. The mathematical formulations for each regime are detailed below.

*Free driving regime (FDR)*: The free driving regime is pertinent when a vehicle is driving without any immediate influence from a leading vehicle, largely because it is positioned at a considerable distance away. Under such conditions, the vehicle's behavior is governed by the following principles:

Acceleration-speed relationship: As the vehicle gains speed, its acceleration tends to decrease. This might be attributed to the natural inclination of drivers to ease off the accelerator as they reach higher speeds.

Tendency towards desired speed: In the free driving regime, the driver will consistently strive to reach and maintain this speed. This desired speed could be based on factors like the vehicle's optimal operating conditions, the driver's personal comfort and safety preferences, or speed limits. Eq. (6) represents the above discussion and formulates the free driving model of our model.

$$\left(\frac{dv}{dt}\right)_{FDR} = a_i \left[1 - \left(\frac{v_i(t)}{v_{oi}}\right)^4\right] \tag{6}$$

*Car-following regime (CFR):* The concept of the car-following regime becomes relevant when vehicles interact along the same path or in the longitudinal direction. Here, it is essential for the following vehicle to respond to the stimuli from the leading vehicle to maintain a safe distance. The assumptions underlying this behavior are:

Acceleration-distance relationship: The acceleration of the following car is directly influenced by the distance to the leading. Specifically, the further away the leading vehicle is, the more the following vehicle tends to accelerate. This is because as the distance increases, there is a natural inclination to close the gap, and vice versa.

Acceleration-speed relationship: Another influencing factor for the acceleration of the following vehicle is the speed of the leading vehicle. As the leading vehicle's speed increases, the acceleration of the following vehicle tends to increase as well. This suggests that the following car tends to match or adapt its speed to that of the leading vehicle, making sure it neither gets too close nor falls too far behind.

Safety gap maintenance: To ensure safety and avoid potential collisions, a minimum distance, referred to as the "bumper-to-bumper gap" or $s_{oi}$, is always maintained between the vehicles. This gap acts as a buffer to allow sufficient reaction time, especially in situations where the leading car brakes suddenly. By understanding and integrating these assumptions, Treiber et al (2000) formulated mathematical car-following behaviors as Eq. (7):

$$\left(\frac{dv}{dt}\right)_{CFR} = a_i \left[1 - \left(\frac{s_i^*(v, \Delta v)}{s_i(t)}\right)^2\right], \tag{7}$$

where

$$s_i^*(v, \Delta v) = s_{oi}(t) + v_i(t) T_i + \frac{v_i(t)\Delta v_i(t)}{2\sqrt{a_i b_i}}. \tag{8}$$

*Behavior adaptation regime (BAR)*: In BAR within our model, we formulate two interconnected concepts: driver behavior adaptation and their risk-taking strategies. These two concepts were integrated while following the three principles of Treiber et al (2013) when enhancing an existing model: (i) minimize the introduction of new parameters, (ii) retain the original interpretation of present parameters, and (iii) maintain simplicity without oversimplifying.

*Formulating driver behavior adaptation in response to task saturation:* Drivers consistently make instantaneous decisions to ensure that the perceived level of driving task saturation remains within their acceptable limit. Drivers maintain their desired level of task saturation based on "task difficulty homeostasis" of Fuller (2011). In essence, drivers are inclined to decelerate (reduce speed or increase the gap between vehicles) when they feel overwhelmed by task saturation. Conversely, they accelerate when they sense that task saturation is below their threshold. Hence, the rate of acceleration in the BAR is inversely proportional to the task saturation. The task saturation, articulated in terms of speed, spacing, and desired time gap (referenced in Eq. (3)), is integral to the BAR of our model. A noteworthy aspect is the drivers' inclination towards a certain deceleration curve when they perceive high task saturation. The smoothness factor $\gamma_i$ modulates this curve (see Eq. (9)). A higher value of $\gamma_i$ implies a more abrupt deceleration (or acceleration) in response to high (or low) task saturation. Conversely, $\gamma_i$ signifies a more gradual deceleration (or acceleration) pattern. In simpler terms, the larger $\gamma_i$ is, the quicker drivers adapt their acceleration as they approach their preferred task saturation level, and vice versa.

*Formulating risk-taking strategy and behavior adaptation*: The degree to which drivers adapt their behavior is also dependent on their individual risk-taking strategies. We postulate that every driver possesses a unique risk-taking strategy when exposed to higher task saturation, symbolized by $\delta_i$ (see Eq. (9)), a parameter that captures risk sensitivity. It is a personal metric reflecting a driver's assessment of driving-associated risks. A higher $\delta_i$ suggests that drivers are more risk-averse and thus more inclined to adapt their behavior when driving task becomes difficult. Conversely, with lower $\delta_i$, drivers are less sensitive to risks and less likely to adapt their response. The parameter spans a range $0 \leq \delta_i < 1$. A comprehensive discussion about the nuances of $\delta_i$ and the underlying risk-taking strategies of drivers is slated for subsequent sections. Building on this discourse, we mathematically articulate the BAR of our model in Eq. (9).

$$\left(\frac{dv}{dt}\right)_{BAR} = a_i \left[1 - \frac{1}{(1-\delta_i)}\left(\frac{v_i(t) T_i}{s_i(t)}\right)^{\gamma_i}\right] \tag{9}$$

Where $\delta_i$ represents an individual driver's unique risk sensitivity parameter, and $\gamma_i$ smoothing factor that shows the drivers how smoothly they decelerate to achieve their desired task saturation. Assembling the three regimes, we have the our IDMTS as Eq. (10).

$$\frac{dv}{dt} = a_i \cdot min\left[\underbrace{1 - \left(\frac{v_i(t)}{v_{oi}}\right)^4}_{Free\ Driving}, \underbrace{1 - \left(\frac{s_i^*(v,\Delta v)}{s_i(t)}\right)^2}_{Car\ Following}, \underbrace{1 - \frac{1}{(1-\delta_i)}\left(\frac{v_i(t) T_i}{s_i(t)}\right)^{\gamma_i}}_{Behavior\ Adaptation}\right] \tag{10}$$

To have the acceleration and deceleration of our model within the boundaries of maximum acceleration and comfortable braking, the third term of our model must satisfy the following conditions:

$$-b_i < a_i \left[1 - \frac{1}{(1-\delta_i)}\left(\frac{v_i(t) T_i}{s_i(t)}\right)^{\gamma_i}\right] < a_i \tag{11}$$

We need to solve the above inequality for values of $\delta_i$. This is of particular importance as values of this parameter near one will lead to instability during model simulation (i.e., harsh braking). Solving the Eq. (11) for values of $\delta_i$ yields Eq (12):

$$\delta_i < 1 \text{ and } \delta_i < \frac{1}{1+\frac{b_i}{a_i}}\left[1 - \left(\frac{v_i(t) T_i}{s_i(t)}\right)^{\gamma_i}\right] \tag{12}$$

Thus, to have the acceleration and deceleration within desirable boundaries, the conditions in Eq. (12) must satisfy. Eq. (12) indicates that values of $\delta_i$ greater than 1 lead to acceleration larger than the maximum acceleration (i.e., $a_i$) and thus should be avoided. On the other hand, the boundary values of $\delta_i$ is subject to other variables variations in Eq. (12).

### 3.4 Illustration of behavioral adaptation and risk-taking strategy with an example of distracted driver

Figure 2 illustrates a visual representation of the driver behavior adaptation, emphasizing the difference between risk aversion and risk-taking strategies. In the first scenario presented in Figure 1 (a), the driver is risk averse and characterized by high risk sensitivity (i.e., high risk sensitivity parameters $\delta$). The driver's journey can be broken down into specific phases.

Moving without interaction phase (time $t_0$): Here, the driver moves effortlessly at their desired speed, free from interaction with the leading vehicle, resulting in an uncomplicated drive that is well within their control. In this instant, the driving behavior is modelled by FDR of IDMTS (i.e., the first term of model).

Merging vehicle in front phase (time $t_1$): A new element is introduced as another vehicle merges directly ahead of the ego driver, inducing a change in dynamics. The short headway with this new vehicle increases task demand on the ego vehicle, though the driving task is still sufficiently within their skill set to manage as the driving task is less than the driver's capacity. In this phase, the interaction with the leading vehicle is controlled by CFR of our model (i.e., the second term of model).

Distraction phase (time $t_2$): The driver receives a phone call. This external distraction acts as an additional layer of task demand to the already evolving driving scenario. The total task demand outweighs the driver's capacity, intensifying the challenge of the driving task. In this phase, the transition from CFR to BAR occurs and the third term of our IDMTS model controls the driving behavior.

Behavior adaptation phase (time $t_2$ to $t_4$): Adopting a risk-averse strategy, the driver adapts their behavior in this situation by reducing speed, naturally increasing the headway, and consequently decreasing the total task demand. This risk aversion strategy is maintained consistently until the phone conversation ends ($t_4$). Thus, the BAR of our model controls the driving behavior in this period until the driver begins the recovery to normal driving behavior.

Back to normal phase (time $t_4$): When the conversation ends, the driver wants to return to a normal situation. In this phase, the transition from BAR to CFR occurs. Therefore, the model shifts from BAR to CFR and the second term of IDMTS controls the driving behavior.

Recovery phase (time $t_5$): The driver wants to recover the lost time by increasing speed. The second term of our model controls the driving behavior.

Conversely, Figure 1 (b) illustrates a risk-taking driver scenario. While the initial conditions mirror the prior scenario, a notable divergence occurs when the driver receives a call at time $t_2$. Unlike the risk-averse driver, this driver does not adapt his (her) behavior due to increased task demand, maintaining their speed and distance from the vehicle ahead. This lack of behavior adaptation makes driving tasks risky during the phone call conversation (from $t_2$ to $t_4$). This choice, coupled with the added distraction, elevates the difficulty and risk during the conversation, signifying a driving strategy marked by lower risk sensitivity (i.e., low risk sensitivity parameter $\delta$).

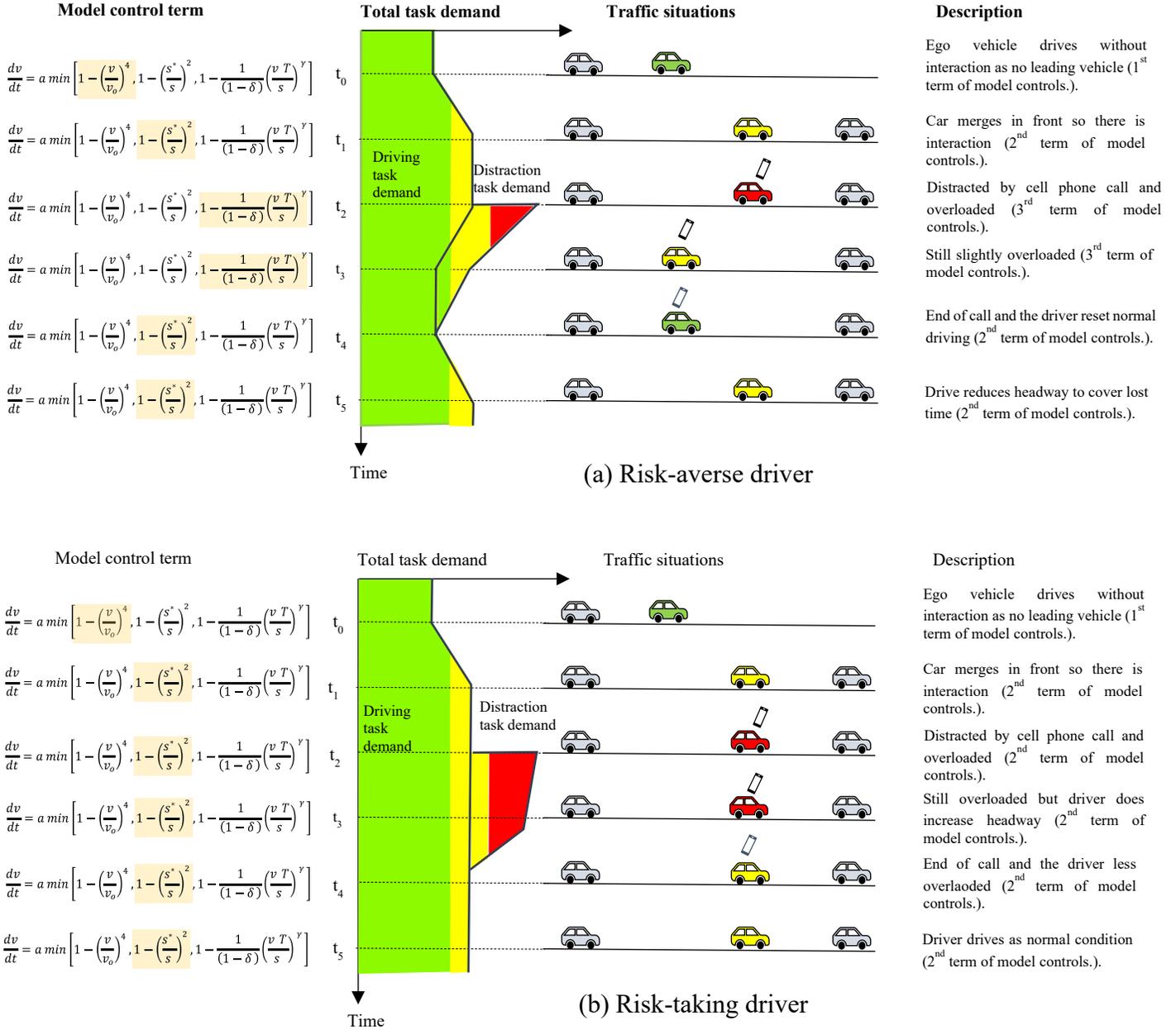

**Fig 1**: Illustration of driver behavioral adaptation and risk-taking strategy in the IDMTS model. Panel (a) shows a risk-averse driver (high risk sensitivity, $\delta$) who adapts headway and speed when task demand exceeds capacity, while panel (b) shows a risk-taking driver (low $\delta$) who sustains higher speed and shorter spacing under distraction. The figure links model control terms, task demand profiles, and traffic situations to highlight the integration of behavioral regimes (FDR, CFR, BAR) within the proposed framework.

### 3.5 Steady state equilibrium and the fundamental diagram of IDMTS

In microscopic models, to describe a stable equilibrium state, it is essential to have uniform driver-vehicle combinations operating on a consistent road environment. This essentially means that all drivers and vehicles are characterized by the same model parameters. Specifically, the acceleration or speed metrics relevant to the model are not distinguished by the type or identity of the vehicle. From a modeling perspective, this steady equilibrium can be defined by two primary conditions that are homogeneous traffic and no accelerations (Treiber et al., 2013). When considering that $\frac{dv}{dt} = 0$ and $\Delta v = 0$ in Eq. (10), it leads to a direct relationship between speed and spacing as Eq. (13).

$$V^e(s) = \begin{cases} \min[v_0, \frac{s-s_0}{T}, (1-\delta)\frac{s}{T}] & s > s_0 \text{ and } \delta < 1 \\ 0 & \text{otherwise} \end{cases} \quad (13)$$

Where is $V^e$ equilibrium speed and the other variables and parameters already described.

From Eq. (13), it is evident that within the context of the FDR, a driver's steady-state equilibrium speed aligns with their desired speed ($v_0$). Conversely, under the CFR, the equilibrium speed is determined not by the driver's preferred speed but by spacing and their desired headway between vehicles. Delving further into BAR scenario, the speed becomes a function of not only the spacing and desired headway, but it is also influenced by the risk sensitivity parameter ($\delta$). Interestingly, the risk sensitivity parameter will affect the equilibrium speed when the risk sensitivity is high (i.e., risk averse drivers) while it will not affect when the risk sensitivity is low (i.e., risk taking drivers). In essence, the degree to which a driver adapts their speed (and/or headway) is contingent upon the magnitude of their risk sensitivity parameter. This highlights the intricate interplay between individual driver characteristics and broader traffic dynamics.

We are keen to delve deeper into the specific traffic conditions and driver trait parameters that influence driver behavior adaptation. Analyzing Eq. (13), it becomes evident that for the BAR of our model — specifically, the third term in the minimization of Eq. (10) to govern the driving behavior (in other words, the behavior adaptation occurs), it needs to be less than FDR and CFR. Consequently, our findings are as follows:

$$v_0 > (1-\delta)\frac{s}{T} \text{ and } \frac{s-s_0}{T} > (1-\delta)\frac{s}{T} \quad (14)$$

Solving Eqs. (14) for $\delta$ to obtain the values of this parameter when it influences the behavior of drivers. We can have the following conditions for $\delta$ for the behavior adaptation. To BAR govern FDR in IDMTS model, the values $\delta$ must satisfy $\delta > 1 - \frac{v_0 T}{s}$. The driver risk sensitivity parameter does have a constant range of values to cause the driver's behavior adaptation. According to this, if the behavior adaptation not only depends on the risk sensitivity but also other parameters ($v_0$ and $T$) and variable ($s$). Further, we can observe that with smaller values of spacing a smaller value of risk sensitivity will lead BAR to govern the FRR in IDMTS model. This is reasonable as smaller values of spacing the driver feels the risk and the driver may no longer be in FDR. To BAR govern the CFR, the $\delta > \frac{s_0}{s}$ must satisfy. This indicates that when the spacing is much larger than the minimum spacing, a smaller value of risk sensitivity will lead BAR to govern the CFR

To establish a micro–macro connection between the micro-level spacing and the macro-level density, we turn to the basic concept of traffic density. By definition, traffic density represents the number of vehicles present over a specific road segment, normalized to the unit length of that segment. Another critical macroscopic parameter is the traffic flow, which fundamentally pertains to the rate of vehicles passing a reference point on the road. A direct method to determine the flow is by considering the inverse of the headway. Bearing the aforementioned relationships in mind and replacing $\rho = 1/s$, $\rho_{max} = 1/s_0$, and $q_{max} = 1/T$ in the Eq. (13), we can have equilibrium speed-density relationships:

$$V^e(\rho) = \begin{cases} \min\left(v_0, q_{max}\left(\frac{1}{\rho} - \frac{1}{\rho_{max}}\right), (1-\delta)\frac{q_{max}}{\rho}\right) & \rho < \rho_{max} \text{ and } \delta < 1, \\ 0 & \text{otherwise} \end{cases} \quad (15)$$

where $\rho$ and $\rho_{max}$ are the density and maximum density, respectively, and $q_{max}$ is the maximum flow. The equilibrium flow can be obtained as $Q^e(\rho) = \rho V^e(\rho)$. Replacing $V^e(\rho)$ from Eq. (15) into this yields Eq. (16) as follows:

$$Q^e(\rho) = \begin{cases} \min\left(v_0\rho, q_{max}\left(1 - \frac{\rho}{\rho_{max}}\right), (1-\delta)q_{max}\right) & \rho < \rho_{max} \text{ and } \delta < 1 \\ 0 & \text{otherwise} \end{cases} \quad (16)$$

Figure 2 presents the three-dimensional equilibrium surfaces of the IDMTS model as a function of density and risk sensitivity. Figure 2(a) shows the flow–density–risk sensitivity surface, which retains the expected unimodal fundamental diagram shape. Increasing risk sensitivity ($\delta$) reduces the maximum attainable flow and shifts the onset of capacity drop, reflecting earlier behavioral adaptation at lower densities. Figure 2(b) depicts the corresponding speed–density–risk sensitivity surface. Speed decreases monotonically with density, with sharper reductions for higher values of $\delta$. Intuitively, these results highlight that greater risk sensitivity (i.e., risk-averse drivers) is associated with lower equilibrium speeds and flows, whereas lower $\delta$ values (i.e., risk-taking drivers) sustain higher flow and speed by tolerating elevated task demands.

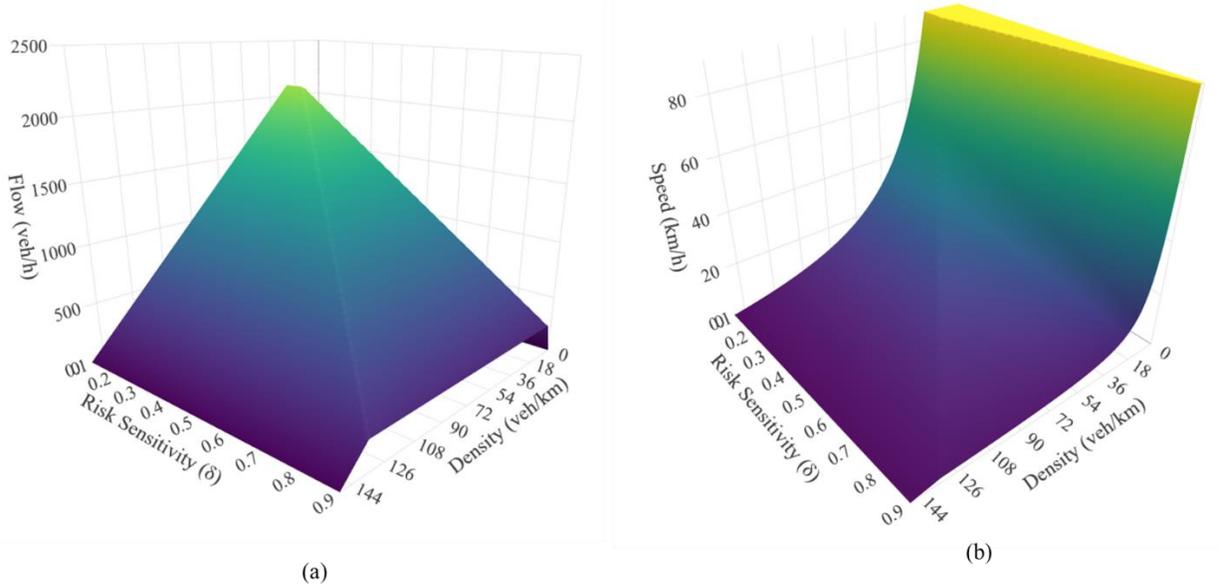

**Fig 2**: Equilibrium surfaces of the IDMTS model under varying densities and risk sensitivities: (a) flow–density–risk sensitivity surface and (b) speed–density–risk sensitivity surface.

To illustrate the transition between regimes and understand the task saturation that triggers regime transition, we simulated the IDMTS model for variations of speed and spacing. Figure 3(a) presents the phase diagram for a risk-averse driver ($\delta = 0.8$), while Figure 3(b) illustrates the same for a risk-taking driver ($\delta = 0.2$). The simulations were conducted using parameter values $a$=1.0 m/$s^2$, $b$=1.5 m/$s^2$, $s_0$=2.0 m, $v_0$=120 km/h, and $T = 1.2$ s, with 2,500 combinations of speed (1–25 m/s) and spacing (10–100 ) tested. Each dot in the figure indicates a simulation experiment. Note, the two divers are considered to have identical driving skills (and thus the same task capacity) and other parameters aside from risk sensitivity. For the risk-averse driver (Figure 3a), the BAR (blue) dominates a larger region of the speed–spacing space, particularly at moderate-to-high speeds and shorter spacings. This indicates that risk-averse drivers adapt more readily under elevated task demand, reducing speed and enlarging spacing to maintain acceptable task saturation. Consequently, the FDR (red) is confined to lower speeds and larger spacings, while the CFR (green) contracts compared to the risk-taking case. Task saturation is generally lower and more evenly distributed, reflecting conservative behaviour and stronger compensatory adaptations.

In contrast, the risk-taking driver (Figure 3b) shows BAR only in a limited portion of the space. Instead, the CFR dominates at moderate spacings, and FDR extends over a wider range of speeds and headways. This behaviour highlights that risk-taking drivers sustain higher speeds and shorter spacings under elevated task demands without significant adaptation, resulting in consistently higher task saturation. Overall, the comparison underscores that higher risk sensitivity induces earlier behavioural adaptation and greater safety margins, while lower risk sensitivity sustains aggressive manoeuvres with elevated task saturation. These differences align with Fuller's Task Capability Interface framework: risk-averse drivers adjust control variables (speed and headway) to reduce overload (Ray Fuller, 2002), while risk-taking drivers tolerate or even ignore overload, accepting narrower safety margins.

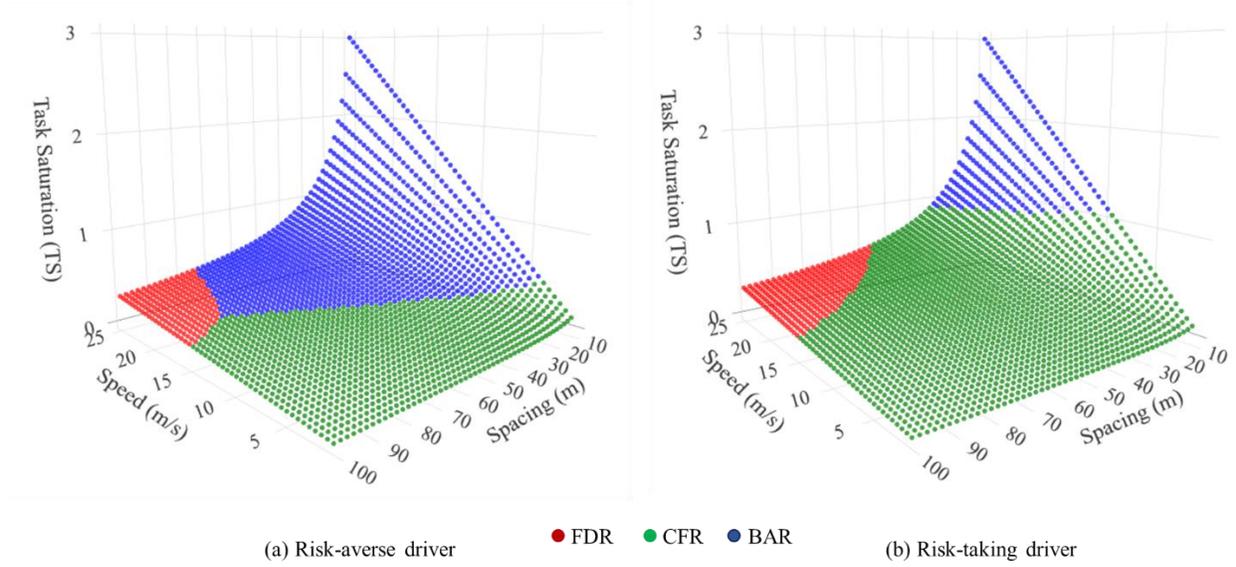

(a) Risk-averse driver     ● FDR    ● CFR    ● BAR     (b) Risk-taking driver

**Fig 3**: Regime phase diagrams of the IDMTS model under varying risk sensitivities: (a) risk-averse driver ($\delta$=0.8) and (b) risk-taking driver ($\delta$=0.2).

### 4. Stability analysis and rational driving Constraints

Since the early development of car-following models, researchers have examined their stability by analyzing how small perturbations evolve over time (local stability) and across a platoon (string stability). This analysis primarily focuses on the progression of a minor disturbance from a leading vehicle both temporally and spatially. Typically, two types of stability are investigated: local linear stability and string linear stability. Local stability concerns the behavior of a vehicle platoon over time when subjected to a small disturbance. A system is considered locally stable if the perturbation does not grow over time, allowing the system to return to equilibrium. String linear stability, on the other hand, deals with the spatial stability of a vehicle platoon under a small disturbance originating from the lead vehicle. A system is deemed string stable if the disturbance decreases progressively across the vehicles; otherwise, it is string unstable (Sun et al., 2020). In this study, we evaluate these two types of stability and discuss the rational driving constraints that a car-following model should possess.

#### 4.1 Local stability and rational driving constraints

Local instability is a particularly severe form of instability, and it is fundamental to design car-following models to avoid it. To analyze the local stability of our proposed car-following model, we consider the steady-state equilibrium flow. Our car-following model is structured as $\dot{v}_i(t) = f(s_i, v_i, \Delta v_i)_t$ and this model is formulated as $f(s_e, v_e, 0)_t = 0$ in the steady state equilibrium condition where $s_e$ and $v_e$ are the equilibrium spacing and speed, respectively. Then, the equilibrium state is perturbed as follows:

$$y_i(t) = s_i(t) + s_e \tag{17}$$
$$u_i(t) = v_i(t) + v_e \tag{18}$$

Where $s_i(t)$ and $v_i(t)$ are small spacing and speed perturbation. The system is stable if the disturbances approach zero over time. For the local stability analysis with small perturbations, we perform a multivariate first-order Taylor expansion of the right-hand side around the equilibrium point, which results in the linearization of the system (further details can be found in (Treiber and Kesting, 2013)). Accordingly, the stability criterion, the local stability criterion of a car-following without time delay, can be expressed as follows:

$$f_v - f_{\Delta v} < 0, \tag{19}$$

where $f_v$ and $f_{\Delta v}$ are the Taylor expansion coefficients of the acceleration function at the steady state, in terms of different variables after first-order linearization. To describe how the acceleration changes with respect to small perturbations in spacing and speed around the equilibrium point, we express these coefficients as follows:

$$f_s = \frac{\partial f}{\partial s}|_e, f_v = \frac{\partial f}{\partial v}|_e, f_{\Delta v} = \frac{\partial f}{\partial \Delta v}|_e \tag{20}$$

In addition to the local stability requirement of a newly developed car-following model, the rational constraints of car-following models are fundamental to ensure the model's dynamics are rational. Accordingly, Wilson et al. (2011) investigated the rational driving constraints of car-following models and concluded that the partial derivatives of sensible car-following models should satisfy the following conditions:

$$f_s > 0, f_{\Delta v} > 0 \text{ and } f_v < 0 \tag{21}$$

Simply put, in two scenarios where all other factors are identical, greater spacing should lead to more acceleration (or less braking). A higher (more positive) relative velocity should result in increased acceleration (less braking) as the leading vehicle is moving away. Additionally, as a vehicle's own velocity increases, the tendency to accelerate should decrease (or the tendency to brake should increase) (Wilson and Ward, 2011).

Theoretically, Eq. (21) should be modified to allow $f_s = 0$ and $f_v = 0$ at larger spacings, where the driving behavior is largely independent of the leading vehicle. To this end, we require the model to behave as follows in conditions where there is no interaction with leading vehicles:

$$f_s = 0 \text{ and } f_v = 0 \tag{22}$$

For reasonable car-following models that satisfy these rational driving constraints that are shown in Eq. (21) and (22), the local stability of such CF models is automatically guaranteed (Wilson and Ward, 2011). To this end, we investigate the rational driving constraints expressed in Eq. (21) and (22) to satisfy both the local stability and rational constraints at the same time. We take partial derivatives of each regime of our car-following model separately. For FDR, Eq. (21) leads to the Taylor expansion coefficients as Eq. (23).

$$(f_s)_{FDR} = 0, (f_v)_{FDR} = -\frac{4av_e^3}{v_0^4}, (f_{\Delta v})_{FDR} = 0 \tag{23}$$

Similarly, for CFR, Taylor expansion coefficients are as follows:

$$(f_s)_{CFR} = \frac{2a}{s_e}\left(\frac{s_0 + Tv_e}{s_e}\right)^2 \tag{24}$$

$$(f_v)_{CFR} = -a\left[\frac{2T(s_0 + Tv_e)}{s_e^2}\right] \tag{25}$$

$$(f_{\Delta v})_{CFR} = \sqrt{\frac{a}{b}} \frac{v_e}{s_e} \frac{s_0 + Tv_e}{s_e} \tag{26}$$

Lastly, the Taylor expansion coefficients for the BAR regime are represented by Eq. (27)-(29).

$$(f_s)_{BAR} = \frac{a\gamma \, v_e T^\gamma}{(1-\delta)} s_e^{-\gamma-1} \tag{27}$$

$$(f_v)_{BAR} = -\frac{a\gamma}{(1-\delta)}\left(\frac{v_e T}{s_e}\right)^{\gamma-1} \frac{T}{s_e} \tag{28}$$

$$(f_{\Delta v})_{BAR} = 0 \tag{29}$$

We would like to evaluate the local stability of each regime. In the FDR, there is no interaction with leading vehicles. Thus, the vehicle relative driving concerning spacing and relative speed must be zero according to Eq. (22). Comparing the stability condition expressed in Eq. (30), the stability condition is expressed as follows:

$$(f_s)_{FDR} = 0, (f_v)_{CFR} = -\frac{4av_e^3}{v_0^4} < 0, (f_{\Delta v})_{FDR} = 0. \tag{30}$$

Eq. (30) meets the stability condition as of Eq. (22). Thus, the stability criteria and rational constraints criteria are met for the FDR. The stability criteria for CFR can be met by Eqs. (31)-(33).

$$(f_s)_{CFR} = \frac{2a}{s_e}\left(\frac{s_0+Tv_e}{s_e}\right)^2 > 0 \tag{31}$$

$$(f_v)_{CFR} = -a\left[\frac{2T(s_0+Tv_e)}{s_e^2}\right] < 0 \tag{32}$$

$$(f_{\Delta v})_{CFR} = \sqrt{\frac{a}{b}}\frac{v_e}{s_e}\frac{s_0+Tv_e}{s_e} > 0 \tag{33}$$

Since all parameters and variables are positive inequalities, one can easily verify that all the above inequalities are true for all values of parameters and variables. Thus, the CFR meet the local stability and rational driving constraints accordingly. Lastly, the BAR local stability and rational driving constraints are expressed as Eqs. (34)-(36).

$$(f_s)_{BAR} = \frac{a\gamma\, v_e T^\gamma}{(1-\delta)} s_e^{-\gamma-1} > 0 \tag{34}$$

$$(f_v)_{BAR} = -\frac{a\gamma}{(1-\delta)}\left(\frac{v_e T}{s_e}\right)^{\gamma-1}\frac{T}{s_e} < 0 \tag{35}$$

$$(f_{\Delta v})_{BAR} = 0 \tag{36}$$

Considering $\delta < 1$, all the inequalities are true for the positive values of parameters and variables. Overall, the proposed car-following model meets the local stability criteria and the rationale driving constraints.

### 4.2 String stability

Stability analysis is a fundamental requirement for evaluating the plausibility of car-following models, as it ensures that small perturbations in speed or spacing do not escalate into unrealistic traffic oscillations. The general stability of car-following models (with no time delay) is indicated in Eq. (37) (Sun et al., 2018).

$$\frac{1}{2} - \frac{f_{\Delta v}}{f_v} - \frac{f_s}{f_v^2} > 0 \tag{37}$$

Following Sun et al. (2018) setup, we evaluated the string stability of each regime in our model. We set the length of the vehicle $l$ = 5 m, minimum gap $s_0$= 2 m, and the desired speed of $v_0$ = 50 km/h and 120 km/h. To observe the string stable and unstable regions, the desired time headway $T$ (0.1-4 s) and maximum acceleration $a$ (0.1-4 m/s²) were varied for CFR. Specifically, we considered 2500 combinations of parameters to observe and visualise stability regions. Similarly, the smoothing factor $\gamma$ (0.1-4) and risk sensitivity $\delta$ (0-0.9) was varied to observe the impact of these parameters on the stability of BAR.

The string stability condition of FDR, following the Eq. (37), leads to $\frac{1}{2} > 0$. The stability criterion is satisfied for all parameters, implying that the system is stable under the given conditions. Thus, the stability criterion for the given values is always met, indicating that this regime is string stable. This is plausible as, in this regime, the vehicle does not interact with other vehicles, so string stability is not an issue in this case. Similarly, the CFR following the Eq. (37), the stability of this regime is as Eq. (38):

$$1 + \frac{v_e}{T}\sqrt{\frac{1}{ab}} - \frac{s_e}{aT^2} > 0 \tag{38}$$

To assess the impact of each parameter, we plotted the stability regions for CFR for various values of desired headway and maximum acceleration (Figure 4). Figure 4 shows that increasing either the maximum acceleration or the desired time headway improves the string stability. Further, higher speed tends to stabilize the system if the other parameters remain constant.

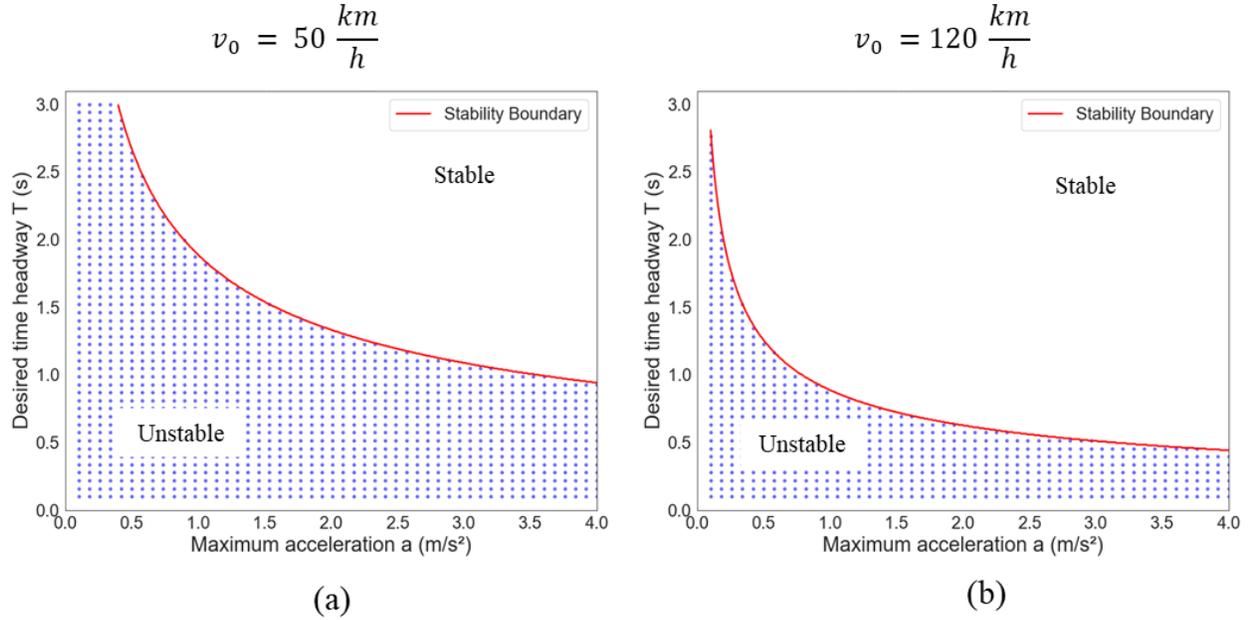

**Fig 4:** String stability region of the IDMTS model in the Car-Following Regime (CFR) for two desired speeds: (a) 50 km/h and (b) 120 km/h. The solid red line separates stable from unstable regions. Each point represents a parameter combination that yields string instability in simulation, while the empty region corresponds to string-stable parameter sets.

Simplifying the string stability condition for BAR leads to Eq. (39). To assess the influence of each parameter on the string stability of this regime, the stability condition for various risk sensitivity $\delta$ and the smoothing factor $\gamma$ are plotted as these two factors play a more significant role in the stability of this regime (Figure 5). According to the figure and Eq. (39), the string stability is improved by higher speed and higher risk sensitivity (risk-taking drivers) values that are plausible. However, these two factors influence stability to a very small extent. The most important parameter for the string stability of this regime is the smoothing factor. Values higher than 1.5 for this factor guarantees the string stability of the regime. This is again plausible as higher values of the factor indicate smoother deceleration that will not destabilize the string of vehicles.

$$\frac{a\gamma v_e^{2\gamma-3} T^\gamma}{(1-\delta)} > 2 \tag{39}$$

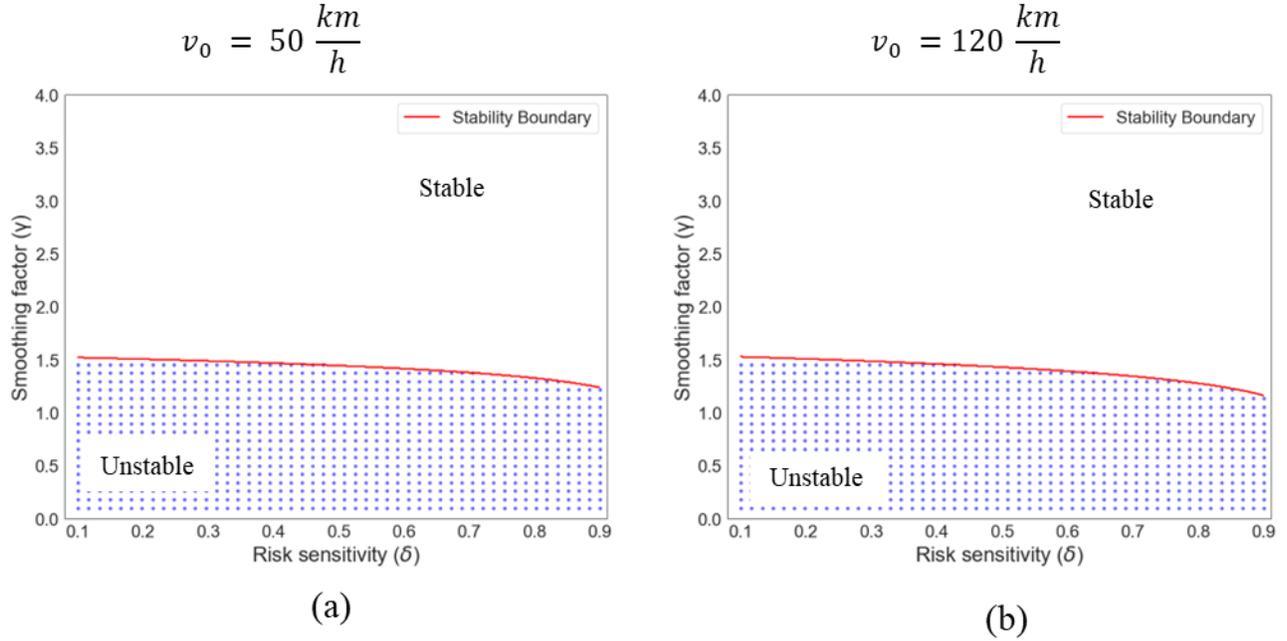

**Fig 5**: String stability region of the IDMTS model in the Behavioural Adaptation Regime (BAR) for two speeds: (a) 50 km/h and (b) 120 km/h. The solid red line, derived from Eq. (37), separates stable and unstable regions. Each point denotes a parameter combination that yields string instability in simulation, while the empty region corresponds to string-stable outcomes as a function of the smoothing factor and risk sensitivity.

## 5. Calibration and Validation

Models are simplified approximations of the real world, which implies that they are subject to errors and limitations (Punzo et al., 2021). Our primary objective is to narrow down these discrepancies and ensure that the model's outcomes align closely with observed values through calibration.

### 5.1 Data

A key methodological choice in this study was to employ two complementary datasets for model calibration and validation: controlled driving simulator experiments and naturalistic trajectory data from the Next Generation Simulation (NGSIM) program (NGSIM, 2016). The motivation for combining these datasets lies in their distinct advantages and limitations. Driving simulator data enables the simultaneous capture of trajectory information and HF inputs such as distraction, which are rarely observable in real-world trajectory datasets. They also allow systematic manipulation of behavioral conditions within a consistent driver population. However, simulator environments may not fully replicate naturalistic driving, particularly with respect to car-following distances and risk-taking tendencies. In contrast, NGSIM I-80 provides high-resolution, real-world trajectory data that reflect authentic traffic conditions and heterogeneous driver behaviors, but it lacks direct HF measurements such as distraction or risk sensitivity. By integrating both sources, the proposed IDMTS model is calibrated and validated against the behavioral richness of simulator data and against the empirical realism of freeway traffic, ensuring both behavioral soundness and practical applicability.

*Driving Simulator Data*: Calibration of our proposed model needs both trajectory data and HF inputs. For this calibration, we utilized trajectory data derived from simulations in which participants were subject to distractions from mobile phone conversations during their drive (Saifuzzaman et al., 2015). This experiment used the CARRS-Q Advanced Driving Simulator, as detailed in the study by Saifuzzaman et al. (2015). The participant group consisted of 32 drivers. Each participant was subjected to two distinct driving scenarios within the simulator: 1) Normal driving without any phone conversation and 2) Driving while engaged in a conversation through a handheld mobile phone.

The focus was on the car-following, which took place on city roads with a designated speed limit of 40 km/h. A 245-meter-long section of the road was chosen to monitor the car-following behaviors of every participant. This particular road stretch comprised four lanes moving in a single direction. Vehicles were parked in outer lanes, restricting the driving to just two middle lanes.

During this car-following event, the participant's vehicle stopped at signal. No intervening vehicles were between the participant's car and these leading ones. As the gap between the participant's vehicle and the foremost vehicle diminished to 60 meters, both leading cars increased their speed to 20 km/h. Upon the gap narrowing to 30 meters or less, these vehicles surged to 35 km/h, maintaining that velocity until the car-following event was completed. The signal at the second intersection remained green to guarantee smooth flow. Essential data observation points, such as the position and velocity of the vehicles, were recorded every 0.05 seconds (Saifuzzaman et al., 2015). From the population of drivers, 22 participants were randomly selected for model calibration, and the rest were used for validation.

*NGSIM I-80 dataset*: To complement the controlled experiments data, we utilised empirical trajectory datasets as well. The dataset captures detailed vehicle movements along a 500-metre segment of the Interstate 80 freeway in Emeryville, California. The dataset records vehicle trajectories at a sampling frequency of 10 Hz, providing high-resolution data on vehicle positions, velocities, accelerations, and headways across varying traffic conditions. For the purpose of our study, we concentrated on lane 2, which exhibited relatively uninterrupted flow with fewer lane changes. The full 45 minutes of data were used, subject to stringent filtering criteria to ensure valid leader–follower pairs: (i) only passenger cars were retained; (ii) each car-following episode was required to last at least 60 seconds; (iii) both leader and follower must remain in the same lane throughout; and (iv) a minimum initial speed differential of 5 m/s was imposed to exclude passive or non-reactive sequences (Ossen and Hoogendoorn, 2011). After filtering, 396 valid car-following pairs were identified. These were randomly split into 277 pairs (70%) for calibration and 119 pairs (30%) for validation.

### 5.2 Measures of Performance and Goodness of Fit

We formulate the calibration problem by defining the model's measures of performance (*MoP*) as a function of the parameter vector $\beta$, as expressed in Eq. (40). The objective is to minimize the deviation between observed and simulated *MoP* using a predefined goodness-of-fit (*GoF*) function, subject to parameter bounds, as shown in Eqs. (41)–(42).

$$MoP^{sim} = F(\beta), \tag{40}$$
$$\min_{\beta} f(MoP^{obs}, MoP^{sim}), \tag{41}$$
$$\text{s.t. } LB_\beta \leq \beta \leq UB_\beta, \tag{42}$$

where $\beta$ denotes the vector of model parameters to be estimated, with lower and upper bounds $UB_\beta$ and $LB_\beta$, respectively. $F(\cdot)$ represents the car-following model under calibration, while $MoP^{obs}$ and $MoP^{sim}$ are the observed and simulated performance measures (e.g., spacing). The function $f(\cdot)$ denotes the *GoF* criterion, which is minimized to obtain the parameter set that best reproduces observed trajectories (Sharma et al., 2019).

To evaluate the performance of models, various *GoF* functions and *MoP* are employed in the literature. Punzo et al. (2021) suggested that when calibrating car-following models, focusing on spacing yields is optimal for spacing and suboptimal results for speed. Conversely, calibrating on speed would result in an optimal speed trajectory but would leave spacing indeterminate. Therefore, spacing is preferred as *MoP* and we adopted this in our study. As for *GoF*, we adopted the preferred Root Mean Squared Error (RMSE) (Punzo et al., 2021), which is defined in Eq. (43). Table 2 shows the range of parameters for calibration of models.

$$RMSE(s) = \sqrt{\frac{1}{T}\sum_{t=1}^{T}(s^{obs}(t) - s^{sim}(t))^2} \tag{43}$$

Where $s^{obs}(t)$ is the observed spacing at time *t*, $s^{sim}(t)$ is the simulated spacing at time *t*, and $T$ is total simulation time.

**Table 2**: Parameters range for calibration of models

| Parameter Description | Parameter (unit) | Minimum | Maximum |
|---|---|---|---|
| Maximum Acceleration | $a$ (m/s$^2$) | 0.50 | 4.00 |
| Desired Deceleration | $b$ (m/s$^2$) | 0.50 | 4.50 |
| Minimum Spacing | $s_0$ (m) | 1.00 | 10.00 |
| Desired Speed | $v_0$ (km/h) | 36.00 | 120.00 |
| Desired Headway | $T$ (s) | 0.20 | 3.00 |
| Risk Sensitivity | $\delta$ | 0.00 | 0.90 |
| Smoothness Factor | $\gamma$ | 1.00 | 4.00 |

### 5.3 Optimization Algorithm

The calibration process seeks to determine the model parameters that reduce the difference between the simulated values and observed data. Generally, this is conducted through optimization techniques (Kashifi, 2024; M T Kashifi, 2024). In this study, we employed the Whale Optimization Algorithm (WOA) to identify the optimum model parameters (Mirjalili and Lewis, 2016). WOA is a population-based metaheuristic inspired by the bubble-net hunting behaviour of humpback whales. The algorithm alternates between exploration (searching for promising regions of the solution space) and exploitation (refining candidate solutions around identified optima) by modelling encircling, spiral updating, and random search mechanisms. This balance enables WOA to avoid premature convergence while maintaining efficient local search capability, making it particularly suitable for nonlinear problems such as car-following model calibration. To ensure consistency and reproducibility, identical optimization settings were applied across all experiments. The WOA was executed with a population size of 200 over 100 iterations, which provided stable convergence behaviour and reliable parameter estimates across repeated runs.

### 6. Results and Discussions

### 6.1 Driving Simulator Data

We calibrated the IDM+ and IDMTS using data from both normal driving and distracted driving scenarios. The results presented in Table 3 indicate the calibrated parameters for both the IDM+ and IDMTS models derived from normal driving trajectory data. The parameters obtained appear to be plausible for both the IDM+ and IDMTS models. Notably, there's a marked similarity between the parameters of IDM+ and IDMTS, which underscores a commendable feature of our proposed model: the introduction of the new cognitive layer adeptly influences driver behavior when deemed necessary, yet it does not perturb the model's original parameters.

**Table 3**: The calibrated parameters of IDM+ and IDMTS with normal driving data. Reported values include mean, standard deviation (std), minimum (min), and maximum (max) across all calibrated trajectories.

|  | IDM+ (Normal driving) | | | | IDMTS (Normal driving) | | | |
|---|---|---|---|---|---|---|---|---|
|  | mean | std | min | max | mean | std | min | max |
| $a$ (m/s$^2$) | 3.45 | 0.90 | 0.88 | 4.00 | 3.55 | 0.68 | 1.06 | 4.00 |
| $b$ (m/s$^2$) | 3.67 | 1.25 | 0.85 | 4.50 | 3.72 | 1.15 | 0.87 | 4.50 |
| $s_0$ (m) | 6.94 | 3.30 | 1.00 | 10.00 | 6.20 | 3.10 | 1.00 | 10.00 |
| $v_0$ (km/h) | 79.49 | 30.37 | 32.45 | 108.00 | 79.06 | 27.20 | 33.93 | 108.00 |
| $T$ (s) | 1.29 | 0.80 | 0.20 | 3.00 | 1.29 | 0.78 | 0.20 | 3.00 |
| $\delta$ |  |  |  |  | 0.59 | 0.30 | 0.04 | 0.90 |
| $\gamma$ |  |  |  |  | 3.00* | 0.92 | 1.00 | 4.00 |
| *RMSE (calibration)* | 2.77 | 0.92 | 0.82 | 4.68 | 2.53 | 0.92 | 0.72 | 4.12 |
| *RMSE (validation)* | 3.63 | 0.98 | 0.93 | 6.32 | 2.63 | 0.94 | 0.76 | 4.72 |

**Note**: * *mode of $\gamma$ as it is considered a categorical variable (values 1, 2, 3, and 4)*

Table 4 outlines the calibrated parameters for both the IDM+ and IDMTS models using distracted driving data. When comparing parameters from normal driving with those from distracted driving, noticeable shifts become apparent. Specifically, for distracted driving scenarios, both maximum acceleration and desirable deceleration have been reduced. This outcome aligns with our expectations: when drivers are occupied in cell phone conversations, they have a higher task demand. Consequently, this heightened cognitive load tends to influence their driving choices, leading them to opt for more conservative acceleration and deceleration values to maintain safety.

In terms of minimum spacing, the IDM+ model indicates a decrease in distracted driving as compared to normal conditions, while the IDMTS model demonstrates only marginal variations. Regarding the desired speed, it's evident that both models (IDM+ and IDMTS) indicate decreased values under distracted conditions relative to normal driving. This is congruent with our assumptions: drivers, particularly those who are more risk-averse, tend to opt for lower speeds under high task demand when their attention is compromised.

Further, we observe that distracted drivers typically choose longer headway than those in normal driving conditions. This finding corroborates our hypothesis: risk-averse drivers, when distracted, prefer extended headway to afford additional decision-making time, acting as a buffer to offset their compromised attention. Lastly, the IDMTS has superior fitting compared to IDM+, which is measured in terms of calibration and validation spacing error.

**Table 4**: Calibrated parameters of IDM+ and IDMTS with distracted driving data

| | IDM+ (Distracted driving) | | | | IDMTS (Distracted driving) | | | |
|---|---|---|---|---|---|---|---|---|
| Parameters | mean | std | min | max | mean | std | min | max |
| $a$ (m/s$^2$) | 3.01 | 1.12 | 0.73 | 4.00 | 3.09 | 1.05 | 0.69 | 4.00 |
| $b$ (m/s$^2$) | 3.24 | 1.44 | 0.50 | 4.50 | 3.73 | 1.08 | 0.90 | 4.50 |
| $s_0$ (m) | 5.97 | 3.20 | 1.00 | 10.00 | 6.13 | 2.77 | 1.00 | 10.00 |
| $v_0$ (km/h) | 78.74 | 30.44 | 27.37 | 108.00 | 77.81 | 26.04 | 27.55 | 108.00 |
| $T$ (s) | 1.68 | 0.93 | 0.20 | 3.00 | 1.55 | 0.85 | 0.20 | 3.00 |
| $\delta$ | | | | | 0.60 | 0.26 | 0.05 | 0.90 |
| $\gamma$ | | | | | 3.00* | 0.98 | 1.00 | 4.00 |
| *RMSE (calibration)* | 2.25 | 1.03 | 0.31 | 4.47 | 1.92 | 1.01 | 0.32 | 4.32 |
| *RMSE (validation)* | 3.21 | 1.02 | 0.39 | 5.32 | 2.32 | 1.03 | 0.36 | 4.68 |

**Note**: * *mode of $\gamma$ as it is considered a categorical variable (values 1, 2, 3, and 4)*

We plotted position, spacing, and speed profiles for three distinct drivers to examine deeper. Figure 7 illustrates the IDM+ and IDMTS simulation for Driver 8, highlighting the driver's adaptation between normal and distracted driving behaviors. In the visual representation, we show three distinct regimes within the IDMTS model (i.e., FDR, CFR and BAR). These are differentiated using color schemes in the spacing profile, aiding in discerning which segment of our proposed model predominantly influences driving behaviors. Upon examination, the IDMTS model consistently exhibits the CFR throughout the period and mirrors the trajectory of the IDM+ model during normal driving conditions for Driver 8. This consistent representation by the IDMTS model is logical, as the TS for this specific driver is minimal in normal driving scenarios. Consequently, the BAR in the IDMTS model does not happen.

However, when observing the spacing profile in distracted driving scenarios, it is evident that the driver has opted for a more extended spacing compared to normal driving. Here, the IDM+ model fails to capture this adapted behavior accurately. This discrepancy manifests as a stark difference between the observed spacing and that predicted by the IDM+ model. In contrast, the IDMTS model aligns more closely with the observed data. Notably, within the distracted driving data, the IDMTS model begins in the CFR state for a brief duration, eventually transitioning to BAR. The initial CFR in the distracted period corroborates our expectation as the spacing is large at the beginning of the car-following, and it decreases at the latter stage, so the task demand is lower at the beginning, and the driver is not oversaturated to adapt their behavior. This pattern contrasts with its behavior during normal driving, where it remained in CFR throughout.

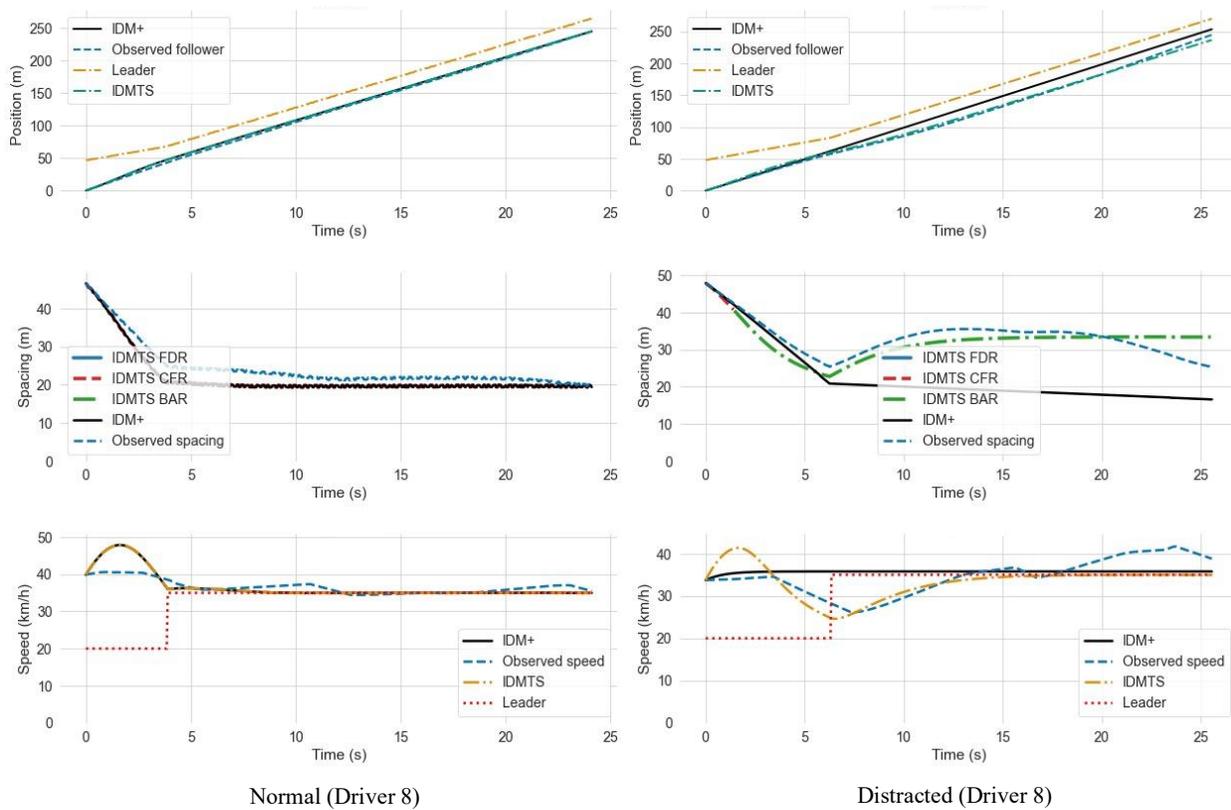

<div align="center">Normal (Driver 8)      Distracted (Driver 8)</div>

**Note**: *IDMTS FDR = IDMTS model in Free Driving Regime, IDMTS CFR = IDMTS model in Car-following Regime, and IDMTS BAR = IDMTS model in Behavior Adaptation Regime*

**Fig 6**: Illustration IDM+ and IDMTS performance in normal and distracted driving condition for risk averse driver 8.

Furthermore, this driver exhibits a risk sensitivity $\delta$ of 0.63 and a smoothness factor ($\gamma$) of 3. The elevated risk sensitivity suggests that this driver is risk-averse. This inclination for risk aversion is further validated by the evident behavior adaptation observed in the data. Concurrently, the $\delta$ value, indicating a less aggressive behavior adaptation, is corroborated by the larger spacing the driver opts for when distracted compared to normal driving.

Figure 8 offers a visual representation of the behavioral traits of Driver 14 as modeled by both the IDM+ and IDMTS models. In the context of normal driving data, the two models depict fairly congruent trajectories for this driver, diverging only slightly. In the beginning, the IDMTS model positions the driver within the CFR for a brief period, subsequently transitioning to the BAR. Interestingly, even though there is no indication of the driver being distracted, their behavior suggests a pronounced risk aversion. This could hint at the driver feeling slightly overwhelmed by the mere act of driving. A closer look reveals the superiority of the IDMTS model in terms of accurately capturing this specific driver's behavior adaptation. It aligns more closely with the observed spacing, whereas the IDM+ model seems to falter in its representation. In the context of distracted driving data, the IDMTS model consistently portrays the driver as operating within the BAR throughout. This portrayal is logical, especially when considering the driver's notably high risk sensitivity, represented by $\delta = 0.88$, and a smoothness factor ($\gamma$) of 3. Such a pronounced risk sensitivity suggests that the driver exhibits strong risk-averse tendencies, prompting them to adapt their behavior even in the face of minor perceived driving risk.

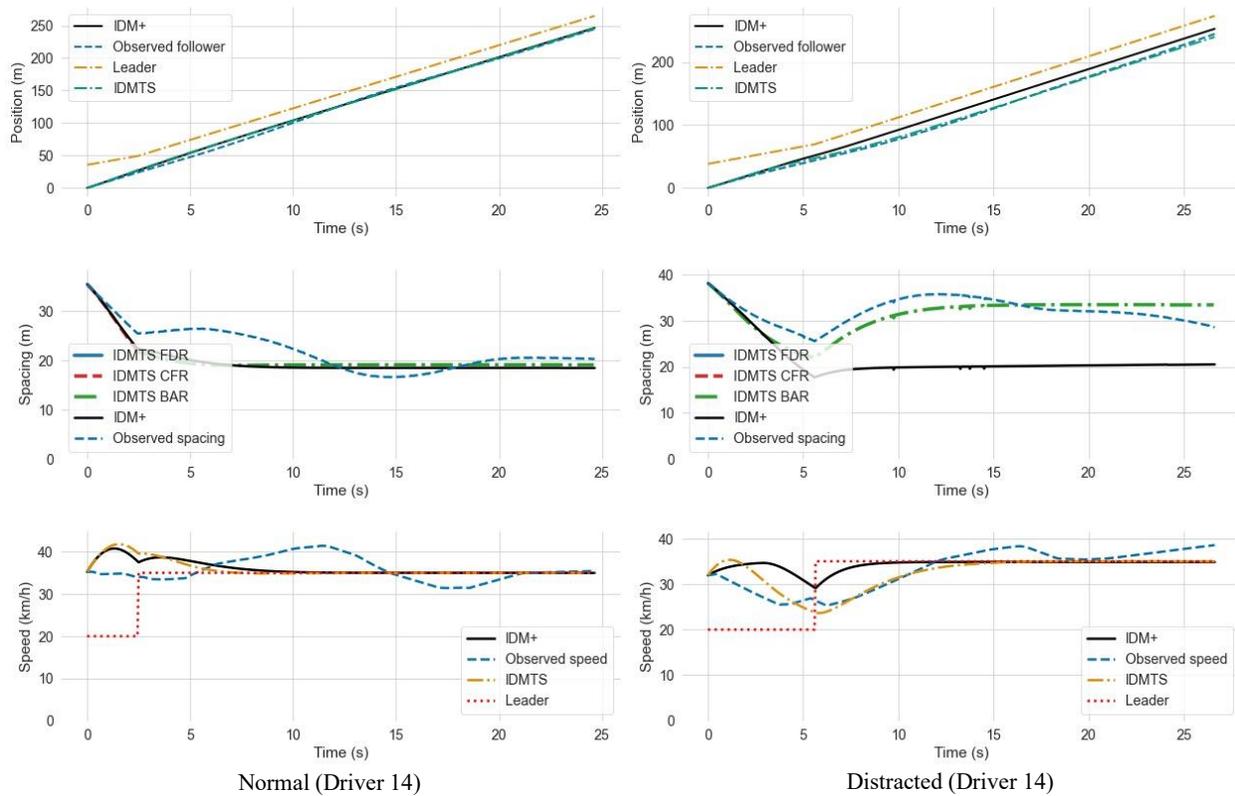

**Fig 7:** Behavior adaptation of IDMTS in distracted driving condition for risk averse driver 14

Figure 9 depicts the behavior of a risk-taking driver, specifically Driver 17. Notably, this driver opts for reduced spacing during distraction compared to normal driving. This is further corroborated by the speed profile, which indicates that Driver 17 tends to select a higher speed when distracted. The calibration and subsequent simulation of the IDMTS model further validate these observations. Within the calibration, this driver exhibits a comparatively lower risk sensitivity value ($\delta = 0.51$). Additionally, the simulation outcomes demonstrate that Driver 17 does not operate within the BAR during distracted intervals. Interestingly, there is a convergence in the IDM+ and IDMTS models in this scenario, primarily because both models function under the CFR, causing our model to essentially mimic the behavior of the IDM+ model.

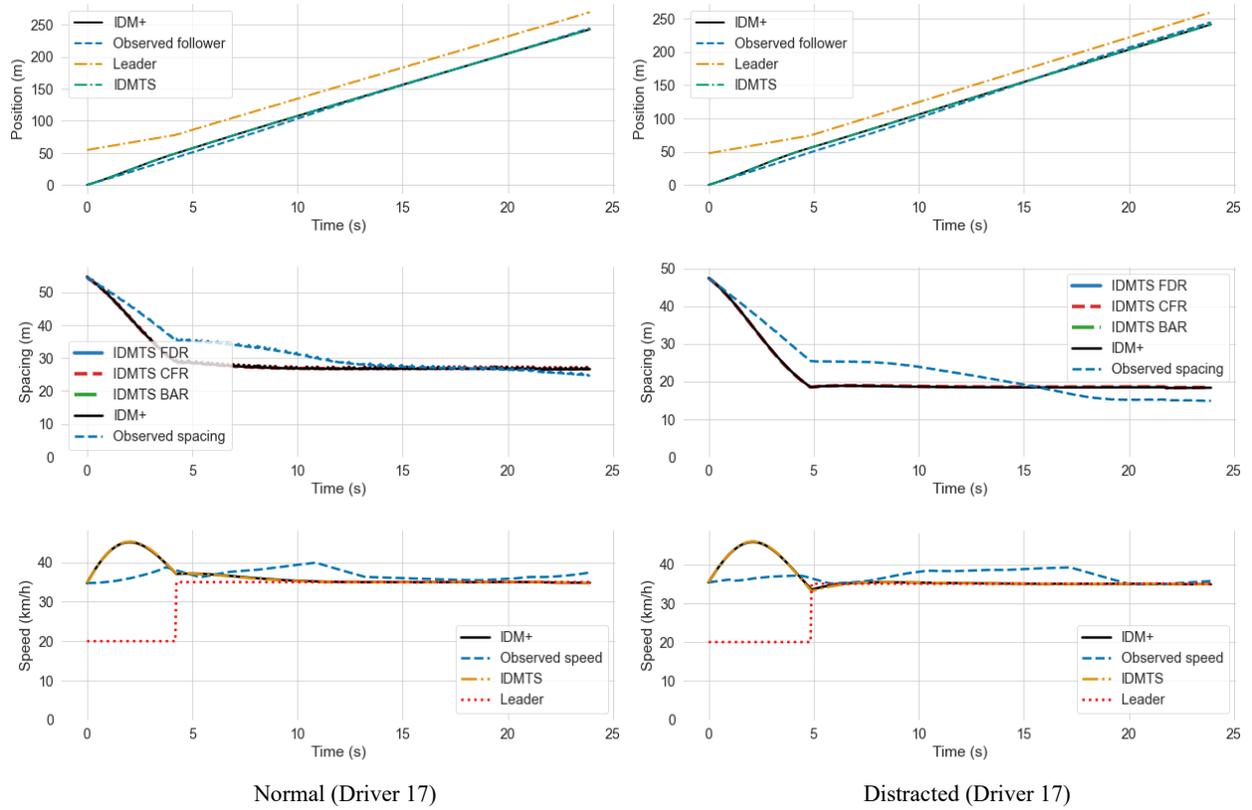

| Normal (Driver 17) | Distracted (Driver 17) |

**Fig 8**: Example of risk-taking driver without behavior adaptation in distracted driving condition

### 6.1 NGSIM Data

Table 5 reports the calibrated parameters of IDM+ and IDMTS on NGSIM I-80. The inclusion of risk sensitivity ($\delta$) and adaptation exponent ($\gamma$) in IDMTS does not significantly disturb the original IDM+ parameters, consistent with the principle suggested by Treiber et al. (2013) that new parameters should extend rather than alter the core model structure. Both models yield reasonable parameters, while IDMTS achieves lower RMSE values in calibration and validation. The additional parameters in IDMTS capture behavioral adaptation and intra-driver risk sensitivity, while preserving the interpretability and stability of the original parameter set.

**Table 5**: Calibrated parameters of IDM+ and IDMTS on NGSIM I-80 data

|  | IDM+ | | | | IDMTS | | | |
| --- | --- | --- | --- | --- | --- | --- | --- | --- |
| Parameters | mean | std | min | max | mean | std | min | max |
| $a$ (m/s$^2$) | 2.16 | 1.23 | 0.73 | 4.00 | 2.31 | 1.15 | 0.72 | 4.00 |
| $b$ (m/s$^2$) | 2.92 | 1.39 | 0.50 | 4.50 | 2.52 | 1.24 | 0.50 | 4.50 |
| $s_0$ (m) | 2.23 | 1.57 | 0.5 | 4.30 | 2.26 | 1.28 | 0.50 | 5.23 |
| $v_0$ (km/h) | 96.76 | 25.17 | 54.16 | 113.99 | 99.46 | 23.39 | 54.87 | 115.54 |
| $T$ (s) | 2.06 | 0.54 | 0.50 | 2.50 | 1.65 | 0.58 | 0.51 | 2.50 |
| $\delta$ | | | | | 0.60 | 0.29 | 0 | 0.9 |
| $\gamma$ | | | | | 3.00* | 1.06 | 1.00 | 4.00 |
| RMSE (calibration) | 4.72 | 1.90 | 1.08 | 12.15 | 3.98 | 2.08 | 0.64 | 12.23 |
| RMSE (validation) | 5.32 | 1.80 | 2.45 | 12.09 | 4.81 | 1.72 | 1.71 | 11.45 |

Figure 10 compares the simulation outputs of the proposed IDMTS model with the baseline IDM+ for two representative leader–follower trajectories. Each column reports position, spacing, and speed profiles, with the IDMTS spacing trajectories further decomposed by regime: Free Driving Regime (FDR, green), Car-Following Regime (CFR, red), and Behavioural Adaptation Regime (BAR, blue). Across both trajectories, the IDMTS model reproduces the observed follower trajectories more closely than IDM+. In the position profiles (top panels), both models follow the leader reasonably well, but IDMTS aligns more consistently with the observed follower, particularly during periods of acceleration and deceleration. The spacing profiles (middle panels) highlight a key distinction: IDM+ tends to overestimate or underestimate spacing in oscillatory regimes, while IDMTS captures regime-specific behaviour. The transitions between FDR, CFR, and BAR are visible in the IDMTS trajectory, with BAR episodes occurring during sharp changes in leader dynamics and CFR dominating steady following. This regime differentiation allows IDMTS to replicate the observed spacing oscillations with greater fidelity. Finally, the speed profiles (bottom panels) show that IDMTS provides closer correspondence with the observed follower speeds, especially in reproducing deceleration phases, whereas IDM+ lags or overshoots during sudden changes. Overall, the results demonstrate that IDMTS extends the predictive capability of IDM+ by embedding regime-specific adaptations, thereby capturing intra-driver behavioural variability and yielding improved agreement with empirical trajectories.

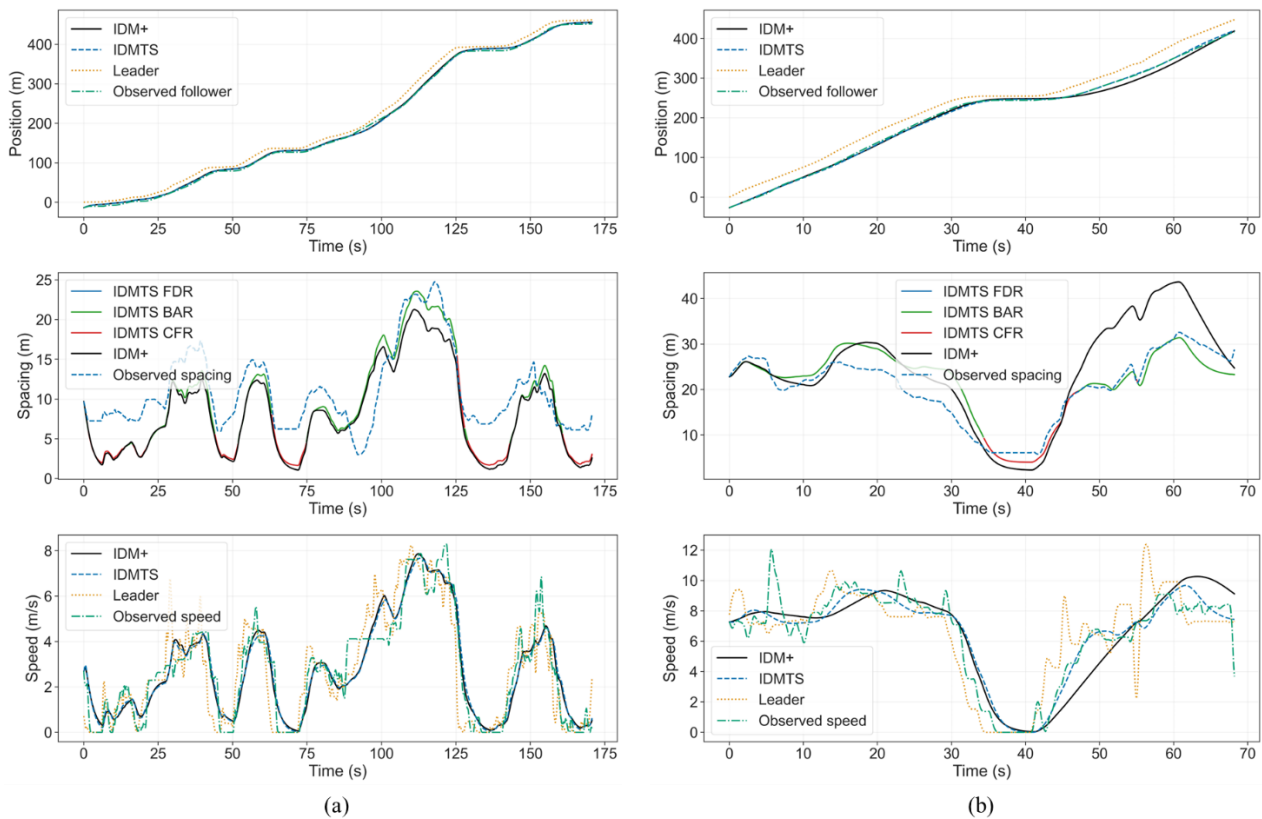

(a)    (b)

**Fig 9**: Simulation for two representative leader–follower pairs comparing IDM+ and IDMTS. In the IDMTS model, spacing profiles are decomposed into regime-specific segments: Free Driving Regime (FDR, green), Car-following Regime (CFR, red), and Behavioral Adaptation Regime (BAR, blue)

## 7. Discussion and Limitations

This study proposed an extended car-following framework that explicitly incorporates behavioral adaptation and task saturation, thereby providing a more comprehensive representation of how human factors influence traffic dynamics. The model demonstrated both numerical and behavioral soundness when tested on driving simulator data and empirical NGSIM I-80 trajectories. In particular, the regime-based formulation allowed us to distinguish between free-driving, car-following, and behavioral adaptation regimes, yielding plausible local and string stability properties.

These results are consistent with the theoretical expectations of stability conditions in human-centered traffic flow models and support the view that incorporating intra-driver variability is critical for reproducing realistic oscillatory patterns.

Despite these advances, genuine limitations should be acknowledged. First, while the proposed parameters $\delta$ and $\gamma$ improved behavioural soundness, they inevitably introduce a degree of model complexity. Although our calibration results indicate that the additional parameters do not compromise generalizability, the risk of over-parameterisation cannot be entirely dismissed, particularly in noisy or short-duration datasets. In such cases, the parameter $\gamma$ can be assumed constant (e.g., $\gamma = 4$) as it will reduce the number of parameters without compromising the behavioural soundness substantially. Second, our experiments focused on longitudinal control in single-lane car-following; future work should examine whether the model generalizes to multi-lane interactions, mixed traffic environments, and connected/automated driving contexts. Third, task saturation was operationalized using a relatively simple functional form. While effective in reproducing observed patterns, more refined specifications of cognitive workload, such as differentiating between visual, cognitive, and manual distractions could provide deeper explanatory power.
Overall, the model strikes a balance between tractability and behavioural richness, offering a framework that can be embedded in both theoretical analysis and simulation practice. Its capacity to integrate human factors into stability analysis makes it a promising tool for evaluating safety-critical situations, policy interventions, and future scenarios involving distracted or risk-sensitive drivers. Nevertheless, the challenge for future research will be to validate the framework across diverse driving populations and real-world contexts, ensuring that behavioural soundness is preserved without compromising computational efficiency in large-scale simulation.

## 8. Conclusion

Over the past decade, the increasing recognition of the importance of human factors (HF) in understanding traffic dynamics has catalyzed significant advancements in traffic flow modeling. This study has contributed to this evolving landscape by introducing the Intelligent Driver Model Task Saturation (IDMTS), a model that synergistically integrates traditional car-following principles with cognitive insights. Drawing upon Fuller's Task Capability Interface (TCI) model, the IDMTS provides a refined representation of human driving behavior, capturing the interplay between behavioral adaptation under high task saturation and individual risk-taking strategies.

Our model was calibrated and validated using data collected from both driving simulators and the NGSIM I-80 dataset. The simulator experiments covered normal and distracted driving scenarios within a consistent driver population, while the trajectory data from NGSIM provided naturalistic evidence of heterogeneous car-following. This dual calibration design ensured that the model not only captured adaptive changes in driver behaviour across contexts but also remained robust to empirical variability observed in large-scale trajectory datasets.

The results demonstrate that the proposed IDMTS model achieves both behavioural soundness and numerical soundness, outperforming IDM+ in reproducing realistic spacing, speed, and adaptation patterns. By explicitly incorporating risk sensitivity and adaptation regimes, IDMTS provides a more faithful representation of intra-driver variability without disturbing the interpretability of the original parameter set. The model's performance across both controlled simulator data and naturalistic field data highlights its versatility and practical relevance for modelling diverse traffic conditions.

A comprehensive stability and rationality analysis further confirmed the plausibility of the proposed framework. The model satisfies rational driving constraints, maintains local stability under realistic parameter ranges, and exhibits string stability characteristics consistent with empirical observations. These results underline that the proposed car-following model is not only theoretically well-founded but also behaviourally credible.

The proposed IDMTS framework is intended to serve two complementary purposes: improving the behavioral realism of car-following models for safety-oriented studies and enhancing the accuracy of traffic operations analyses. To support these applications, calibration should be undertaken using trajectory data that reflects a representative spectrum of drivers within the target population. This may be achieved through naturalistic trajectory datasets, instrumented vehicle studies, or controlled driving simulator experiments. Since IDMTS is formulated as a general extension that captures risk sensitivity and regime-specific adaptation rather than being tied to behavioral factors, it can accommodate a wide range of driver heterogeneities and task demands. The calibrated model can thus be employed both as a simulation tool, comparable to conventional car-following models but with enhanced behavioral soundness, and as an evaluation tool to quantify the implications of risk sensitivity and adaptation regimes on traffic safety and operational performance by varying risk sensitivity.

As the traffic flow theory community continues to evolve, this study lays the groundwork for future research to explore the involved dynamics between HF and traffic flow dynamics. By advancing our understanding of how human behavior shapes traffic dynamics, we can better inform traffic management strategies, enhance safety measures, and ultimately contribute to more efficient transportation systems. Future studies should seek to validate our model in additional dimensions of human behavior, such as the effects of perception and response errors, to further refine our understanding of driver dynamics. Ultimately, this research not only enriches theoretical discourse but also has practical implications for improving road safety and traffic efficiency in increasingly complex driving environments.

**Acknowledgements**

## 9. References

Fuller, Ray, 2002. Psychology and the highway engineer. Fuller & Santos.
Fuller, R, 2002. Human factors and driving. Human factors for highway engineers.
Fuller, R., 2005. Towards a general theory of driver behaviour. Accid Anal Prev 37, 461–472.
Fuller, R., 2011. Driver control theory: From task difficulty homeostasis to risk allostasis. In: Handbook of Traffic Psychology. Elsevier, pp. 13–26.
Gipps, P.G., 1981. A behavioural car-following model for computer simulation. Transportation Research Part B: Methodological 15, 105–111.
Hamdar, S.H., Mahmassani, H.S., Treiber, M., 2015. From behavioral psychology to acceleration modeling: Calibration, validation, and exploration of drivers' cognitive and safety parameters in a risk-taking environment. Transportation Research Part B: Methodological 78, 32–53.
Heino, A., van der Molen, H.H., Wilde, G.J.S., 1996. Differences in risk experience between sensation avoiders and sensation seekers. Pers Individ Dif 20, 71–79.
Hoogendoorn, R., Van Arem, B., Hoogendoorn, S., 2013. Incorporating driver distraction in car-following models: Applying the TCI to the IDM. In: 16th International IEEE Conference on Intelligent Transportation Systems (ITSC 2013). IEEE, pp. 2274–2279.
Information, I., 2016. Next Generation Simulation ( NGSIM ) Interstate 80 Freeway Dataset Identification Information Time Period of Content.
Kashifi, Mohammad Tamim, 2024. Robust spatiotemporal crash risk prediction with gated recurrent convolution network and interpretable insights from SHapley additive explanations. Eng Appl Artif Intell 127, 107379.
Kashifi, M T, 2024. Deep Hybrid Attention Framework for Road Crash Emergency Response Management. IEEE Transactions on Intelligent Transportation Systems 1–12.
Kochi, F., Saito, Y., Uchida, N., Itoh, M., 2023. Task difficulty, risk feeling, and safety margin in the determination of driver behavior to prepare for traffic conflicts. Accid Anal Prev 192, 107284.
Lewis-Evans, B., De Waard, D., Brookhuis, K.A., 2010. That's close enough—A threshold effect of time headway on the experience of risk, task difficulty, effort, and comfort. Accid Anal Prev 42, 1926–1933.
May, A.D., 1990. Traffic flow fundamentals.
Mirjalili, S., Lewis, A., 2016. The Whale Optimization Algorithm. Advances in Engineering Software 95, 51–67.
Ossen, S., Hoogendoorn, S.P., 2011. Heterogeneity in car-following behavior: Theory and empirics. Transp Res Part C Emerg Technol 19, 182–195.
Pipes, L.A., 1953. An operational analysis of traffic dynamics. J Appl Phys 24, 274–281.
Punzo, V., Zheng, Z., Montanino, M., 2021. About calibration of car-following dynamics of automated and human-driven vehicles : Methodology , guidelines and codes. Transportation Research Part C 128, 103165.
Saifuzzaman, M., Zheng, Z., 2014. Incorporating human-factors in car-following models: A review of recent developments and research needs. Transp Res Part C Emerg Technol 48, 379–403.
Saifuzzaman, M., Zheng, Z., Haque, M.M., Washington, S., 2015. Revisiting the Task–Capability Interface model for incorporating human factors into car-following models. Transportation research part B: methodological 82, 1–19.

Saifuzzaman, M., Zheng, Z., Haque, M.M., Washington, S., 2017. Understanding the mechanism of traffic hysteresis and traffic oscillations through the change in task difficulty level. Transportation Research Part B: Methodological 105, 523–538.

Schakel, W.J., Knoop, V.L., Van Arem, B., 2012. Integrated lane change model with relaxation and synchronization. Transp Res Rec 2316, 47–57.

Sharma, A., Zheng, Z., Bhaskar, A., 2019. Is more always better? The impact of vehicular trajectory completeness on car-following model calibration and validation. Transportation research part B: methodological 120, 49–75.

Sun, Jie, Zheng, Z., Sun, Jian, 2018. Stability analysis methods and their applicability to car-following models in conventional and connected environments. Transportation Research Part B: Methodological 109, 212–237.

Sun, Jie, Zheng, Z., Sun, Jian, 2020. The relationship between car following string instability and traffic oscillations in finite-sized platoons and its use in easing congestion via connected and automated vehicles with IDM based controller. Transportation Research Part B: Methodological 142, 58–83.

Treiber, M., Hennecke, A., Helbing, D., 2000. Congested traffic states in empirical observations and microscopic simulations. Phys Rev E 62, 1805.

Treiber, M., Kesting, A., 2013. Traffic flow dynamics. Traffic Flow Dynamics: Data, Models and Simulation, Springer-Verlag Berlin Heidelberg 983–1000.

Treiber, M., Kesting, A., Treiber, M., Kesting, A., 2013. Elementary car-following models. Traffic Flow Dynamics: Data, Models and Simulation 157–180.

van Lint, J.W.C., Calvert, S.C., 2018. A generic multi-level framework for microscopic traffic simulation—Theory and an example case in modelling driver distraction. Transportation Research Part B: Methodological 117, 63–86.

Van Lint, J.W.C., Calvert, S.C., 2018. A generic multi-level framework for microscopic traffic simulation—Theory and an example case in modelling driver distraction. Transportation Research Part B: Methodological 117, 63–86.

Wiedemann, R., 1974. Simulation des Strassenverkehrsflusses.

Wilson, R.E., Ward, J.A., 2011. Car-following models: fifty years of linear stability analysis–a mathematical perspective. Transportation Planning and Technology 34, 3–18.

Yang, H.-H., Peng, H., 2010. Development of an errorable car-following driver model. Vehicle System Dynamics 48, 751–773.

Zhu, M., Wang, X., Tarko, A., Fang, S., 2018. Modeling car-following behavior on urban expressways in Shanghai: A naturalistic driving study, Transportation Research Part C: Emerging Technologies.